\documentclass[a4paper, final]{article}   

\usepackage[margin=1.5in]{geometry}
\usepackage[latin1]{inputenc}  %for special German letters
\usepackage[T1]{fontenc}
\usepackage[ngerman,english]{babel}
\usepackage[english=american]{csquotes}

\usepackage{graphicx}

\usepackage{amsmath}
\usepackage{amsthm}
\usepackage{amssymb}
\usepackage{cleveref}

\usepackage{siunitx} %for units \SI{1}{\micrometre} and numbers \num[decimalsymbol=comma]{12,34}
\usepackage[backend=biber, sorting=none, maxbibnames=3, doi=true]{biblatex}
\bibliography{DynamicallyDilutedAlignmentModel_Rxiv.bib}

\usepackage{authblk}  %for title page with authors and affiliations
\newcommand{\define}{\mathrel{\mathop:}=}
\newcommand{\enifed}{=\mathrel{\mathop:}}
\newcommand{\transpose}{T}
\newcommand{\MFA}[1]{\hat {#1}}   %symbol for mean-field approximation
\newcommand{\linMFA}[1]{\bar{#1}} %symbol for linearised mean-field approximation linearisation
\newcommand{\tmo}{T_{mo}}         %symbol for time of minimal order in IPS
\newcommand{\tmoMF}{\hat{T}_{mo}} %symbol for time of minimal order in mean-field approximation
\newcommand{\tmolinMF}{\overline{T}_{mo}}  %symbol for time of minimal order in linear mean-field approximation

\title{A Dynamically Diluted Alignment Model Reveals the Impact of Cell Turnover on the Plasticity of Tissue Polarity Patterns}

\author[1,2]{Karl B.~Hoffmann}
\author[1,3]{Anja Voss--B{\"o}hme}  %Anja wants the {\ss} to be written as plain  ss
\author[4]{Jochen C.~Rink}
\author[1,5,6]{Lutz Brusch}
\affil[1]{Center for Information Services and High Performance Computing, Technische Universit{\"a}t Dresden, Dresden, Germany}
\affil[2]{now at Faculty of Computer Science, Technische Universit{\"a}t Dresden, Dresden, Germany, at Max Planck Institute of Molecular Cell Biology and Genetics, Dresden, Germany, and at Center for Systems Biology Dresden, Dresden, Germany}
\affil[3]{University of Applied Sciences Dresden, Dresden, Germany}
\affil[4]{Max Planck Institute of Molecular Cell Biology and Genetics, Dresden, Germany}
\affil[5]{Center for Advancing Electronics Dresden, Technische Universit{\"a}t Dresden, Dresden, Germany}
\affil[6]{To whom correspondence should be addressed. Email: {lutz.brusch{@}tu-dresden.de} }

\date{}

\begin{document}

\maketitle

\begin{center}
Preprint as of June 23, 2017, prior to submission to \emph{Journal of the Royal Society Interface}.
\end{center}

\begin{abstract}
The polarisation of cells and tissues is fundamental for tissue morphogenesis during biological development and regeneration. 
A deeper understanding of biological polarity pattern formation can be gained from the consideration of pattern reorganisation in response to an opposing instructive cue, which we here consider by example of experimentally inducible body axis inversions in planarian flatworms.  
Our dynamically diluted alignment model represents three processes: entrainment of cell polarity by a global signal, local cell-cell coupling aligning polarity among neighbours and cell turnover inserting initially unpolarised cells. 
We show that a persistent global orienting signal determines the final mean polarity orientation in this stochastic model. 
Combining numerical and analytical approaches, we find that neighbour coupling retards polarity pattern reorganisation, whereas cell turnover accelerates it. 
We derive a formula for an effective neighbour coupling strength integrating both effects and find that the time of polarity reorganisation depends linearly on this effective parameter and no abrupt transitions are observed. 
This allows to determine neighbour coupling strengths from experimental observations. 
Our model is related to a dynamic $8$-Potts model with annealed site-dilution and makes testable predictions regarding the polarisation of dynamic systems, such as the planarian epithelium.
\\ 

\end{abstract}

\section{Introduction}
%
% 1,2. Apical-basal
Epithelial tissues can be considered as two-dimensional sheets of densely packed cells. 
The properties of epithelia are highly regulated and instrumental for morphogenesis during biological development and regeneration. 
A key property of epithelia is the establishment of different membrane domains on either side of the plane, termed apical and basal. 
This process polarises epithelial cells perpendicular to the  plane \cite{Bryant2008}. 
Molecules responsible for the establishment of apico-basal polarity include phosphoinositides, various GTPases, and the Crumbs and PAR complexes \cite{Rodriguez2014}. 

% 3. PCP
By asymmetrically localizing an independent set of molecules including Frizzled/ Flamingo and Fat/Dachsous along an axis perpendicular to the apical-basal axis, cells of many epithelia superimpose a second polarity pattern \emph{within} the plane, termed planar cell polarity (PCP) \cite{GoodrichStrutt2011}.
PCP controls fundamental processes during embryonic development and tissue regeneration in many species including actin filament orientation, convergence-extension, tissue reshaping, sensory organ formation, wing hair orientation, directional tissue growth and animal locomotion \cite{GoodrichStrutt2011,LawrenceCasal2013,SebbaghBorg2014,Axelrod2009,RodrigoAlbors2015,Seifert2007,Viktorinova2011}.

% 4. Two synergistic PCP mechanisms in principle
Mechanistically, PCP and the resulting planar tissue polarity integrate two general classes of inputs.
(1) Global cues provided by the slope of tissue-scale gradients. These can consist of ligand concentration profiles \cite{Gao2011,Wu2013}, gene expression gradients \cite{Adler1997}, or mechanical shear stress \cite{AigouyEtAlEaton2010}.
(2) Local cues provided by cell-cell coupling.  The alignment of cell polarisation vectors among neighbouring cells propagates anisotropies from tissue boundaries or mutant clones and is mediated by the differential distribution of PCP and/or Fat/Dachsous components across cell/cell interfaces \cite{Lawrence2007,GoodrichStrutt2011,Ambegaonkar2012,Sagner2012}.
These mechanisms are universally found across many species and tissues.
In most contexts, both inputs act synergistically to establish and maintain planar tissue polarity \cite{SimonsMlodzik2008,Peng2012}.

% 5. Models of PCP and understood behaviour
Theoretical studies of the collective phenomena of PCP confirmed that cell-cell neighbour coupling fosters a uniform polarity response of all cells to noisy and non-monotonous tissue-scale signals \cite{AmonlirdvimanKhareEtAl2005,LeGarrecLopezKerszberg2006,Wang2006a,MaAmonlirdviman2008,Zhu2009,BurakShraiman2009,Viktorinova2011,Fischer2013,Zhu2013,ManiGoyalIrvineShraiman2013}.
In particular, weak and even transient biases stemming from a polarised boundary or graded signal suffice to orient an entire epithelium when present from the \emph{onset} of PCP dynamics in initially unpolarised cells \cites{BurakShraiman2009}.
% 5b. Potts model of ferromagnetism, asymptotic state
Understanding of the underlying principle can be gained from statistical physics:
The $q$-Potts model studies two-dimensional lattices that allow $q \in \mathbb{N}$ discrete polarisation vectors in $\mathbb{R}^2$ \cite{Wu1982}.

Indeed, the emergence of long-range order in PCP bears analogy to ferromagnetism. 
In the above models, 
each lattice node (or cell) carries a vectorial magnetic moment, analogous to a cell's PCP vector.
That system's energy decreases by favouring configurations where individual magnetic moments align among neighbours and with the vector of an external magnetic field \cite{Heisenberg1928}.
When fluctuations that tend to randomise individual magnetic moments are below a critical value, then long-range order and a system-wide net magnetisation emerge spontaneously also in the absence of an external bias \cite{Landau1980}.
Analogously, PCP patterns in mutant tissue of fly wings and in model simulations, that abolished or decoupled the external bias, show spontaneously emerging order \cite{BurakShraiman2009}.

Contrary to the fixed arrangement of spins in ferromagnetic matter, however, biological tissues are composed of living cells that are born, age and become eliminated from the tissue. 
Tissues often exist much longer than their constituting individual cells and many tissues maintain their polarised state despite continuous cell turnover. 
In general, there are two scenarios how new cells can establish their PCP, either inherit PCP from their polarised mother cells or polarise de-novo. 
Since PCP signalling depends on the state of neighbouring cells, such cell turnover not only modulates the PCP state locally but constitutes a topological perturbation of the cell arrangement, modulating the number of signalling neighbours. 
This may fundamentally alter the system dynamics beyond that of the classical Potts model.

% 6. heterogeneity (diluted Potts model)
Diluted variants of the Potts model with zero magnetic moments for a subset of nodes, where the zero nodes are either fixed ("quenched site-dilution") or in thermodynamic equilibrium with the other states ("annealed site-dilution"), have been studied for the equilibrium distribution and properties of the asymptotic state \cite{Wu1982}.
However, less is known about the duration and trajectory of transient dynamics approaching the asymptotic state, and on the impact of site-dilution on them. 
This requires to model the process of cell turnover directly according to a specific experimental system.
%This variant, however, does not capture the dynamic nature of cell turnover.

% 7. Open question: transient dynamics and duration
Our work has two objectives.
First, it shall bridge the gap between the existing models with/without static site-dilution and the dynamics of polarity in tissues with cell turnover.
We propose a dynamic model, similar to an $8$-Potts model with annealed site dilution, termed \textit{dynamically diluted alignment model} in the following, for the study of planar polarity formation and maintenance in biological tissues.
Second, it shall elucidate the transient dynamics approaching the asymptotic state. 
We propose that new insight into polarity pattern formation can be gained from analysing the particular transient dynamics of polarity reorganisation when an initially coherent polarity pattern is confronted with an opposing instructive signal.
We therefore ask, in which way the contradiction between inputs is resolved and how the time requirement for conflict resolution depends on parameters, especially the cell birth and death rates.

\begin{figure}

\def\svgwidth{1.0 \textwidth}
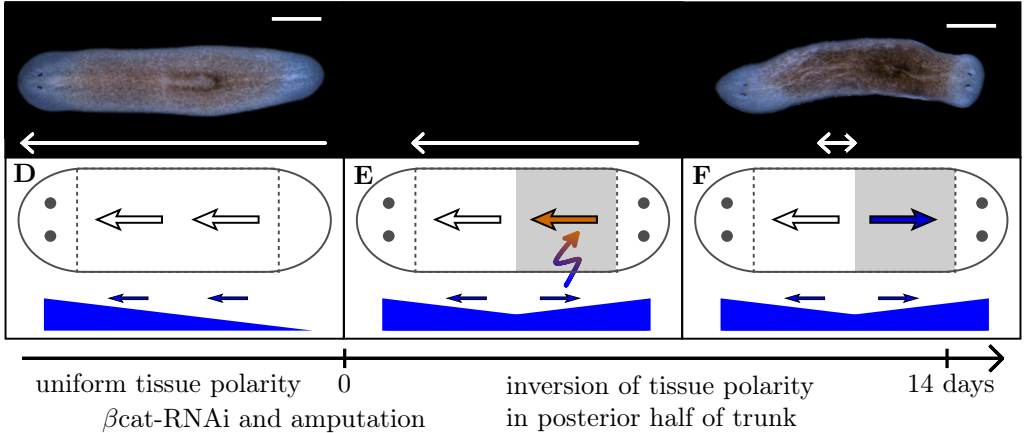

\caption{
Dynamic inversion of pre-existing planar tissue polarity.
\textbf{A-C.} Planaria \textit{Schmidtea mediterranea}, anterior left, scale bar~\SI{500}{\micro \metre}. Arrows indicate gliding direction and speed.
\textbf{A.} Normally regenerated animal. 
\textbf{B-C.} Smed-$\beta$-catenin-1-RNAi treated animals after head and tail amputation (B., 3 days post amputation) regenerate heads at both ends (C., 14 days post amputation, arrowheads indicate eyespots of abnormally regenerated head). 
\textbf{D-F.} Sketches of locally predominating direction of motion (thick arrows) as a proxy for the planar tissue polarity pattern. 
Blue profiles below indicate hypothetical long range signals from the nearest head and the resulting gradient vector (thin arrows). 
\textbf{E.} Polarity of the posterior trunk is locally coherent but conflicts with gradient direction (lightning arrow). 
Our model focuses on that grey shaded region.
\textbf{F.} Symmetric tissue polarity in coherence with local neighbourhood and gradient direction. 
}
\label{fig:exp_InvertedMotionDirectionIndicatesPolarityInversion}
\end{figure}

% 8. Study situation with conflict in Planaria
The biological inspiration for our approach is the experimentally inducible inversion of global body plan polarity in the planarian \textit{Schmidtea mediterranea} \cites{GurleyRinkAlvarado2008}{PetersenReddien2008}. 
The regeneration of a second head instead of a tail (see fig.~\ref{fig:exp_InvertedMotionDirectionIndicatesPolarityInversion}A-C) can be assumed to constitute a conflicting cue for the polarisation pattern in pre-existing tissues. 
The multi-ciliated ventral epithelium is likely to be one such planarly polarised tissue \cite{Wallingford2010,Azimzadeh2016}. 
Its cilia drive the gliding locomotion of planarians, implying consistent polarisation of individual cilia and thus of the ventral epithelium as a whole \cite{Rink2009,Rompolas2013}. 
Consequently, the movement of the animals may also inform on polarisation phenomena within the epithelium. 
In experimentally generated double-headed animals, each of the two heads moves into opposite directions, thus giving rise to a continuous tug-of-war between the two heads with little net movement but stretching and thereby thinning the bulk tissue \cite{GurleyRinkAlvarado2008}.
We interpret the balanced bi-polar movement as evidence for a re-polarisation of the pre-existing epithelium (gray area in fig.~\ref{fig:exp_InvertedMotionDirectionIndicatesPolarityInversion}E,F) in response to an instructive cue provided by the new head. 
Moreover, we hypothesize that the cue constitutes a gradient of a signalling molecule, analogous to the Wnt gradient that patterns the planarian tail \cite{AdellCebriaSalo2010,Almuedo2011,Stueckemann2017}.
This interpretation is further supported by the observed symmetric inward motion of both body halves in double-tailed planaria \cites[movie~S3]{GurleyRinkAlvarado2008}[movie~S2]{RinkGurleyElliottAlvarado2009}.

% 9b. Cell turnover from neoblasts
Our work explores the question of how the polarisation of a cell field responds to the superposition of a conflicting long-range signal. 
We assume that constituent cells undergo continuous turn-over via the integration of new, initially unpolarised cells born outside the tissue (the equivalent of planarian neoblast progeny  \cites{Alvarado2006}{Rink2013}) and the balanced extrusion of old, polarised cells in a dynamic steady state. 
A transiently na{\"i}ve cell presents no polarity information to its neighbours, and with a certain rate turns into a polarised state itself \cite{Chung2014}. Such dynamic loss and re-establishment of polarisation is here modelled as dynamic (annealed) site dilution of an $8$-Potts model.

% 10. Paper outline
This article has the following structure.
We first develop a dynamically diluted alignment model in the framework of Interacting Particle Systems which accounts for a global orienting signal, local coupling, and cell turnover.
This dynamically diluted alignment model allows to study the effects of cell turnover on polarity patterns.
Specifically, we ask whether and how polarity patterns with coherent initial polarisation counter-directional to the global signal reorganise, to resolve the conflict between local and global directional cues, and what the time requirement is if they do so.
We consider a polar alignment order parameter and identify the corresponding time of minimal order as the key characteristic of transient dynamics.
Simulating the full model and by theoretical as well as numerical analysis of a mean-field approximation, we then show that cell-cell neighbour coupling in addition to its synergistic and noise-filtering role mentioned above \emph{retards} the response of planar tissue polarity to dynamically changing global inputs whereas cell turnover \emph{accelerates} it.
Finally, we establish a relation of the system parameters that determines the time requirement for polarity reorganisation.
We close with a discussion of these results.

%============================================================================================================================================
\section{Mathematical Model of Cell Polarity and Turnover} %\section{Methods}

\subsection{Model definition} \label{subsec:Model_Def}

%0 Model structure
We define an Interacting Particle System (IPS)~\cites{Liggett1985}{IPS_in_EncyclopediaSystemsBiology}{KlaussVoss-Boehme2005}{KlaussVoss-Boehme2008} model for tissue polarity dynamics at the cellular level, that incorporates polarity alignment with respect to a global signal and to neighbours' polarity vectors as well as cell turnover.
The model cells occupy the nodes of a finite two-dimensional square lattice $S$ that represents the epithelial tissue subjected to initially conflicting signals, as for instance the grey-shaded area in fig.~\ref{fig:exp_InvertedMotionDirectionIndicatesPolarityInversion}E,F.
The cellular scale of granularity allows to describe the essential interactions yet keeps the model analytically tractable, in analogy to the variants of the Potts model studying ferromagnetism.

\begin{figure}
\includegraphics[width=1.0 \textwidth, keepaspectratio]{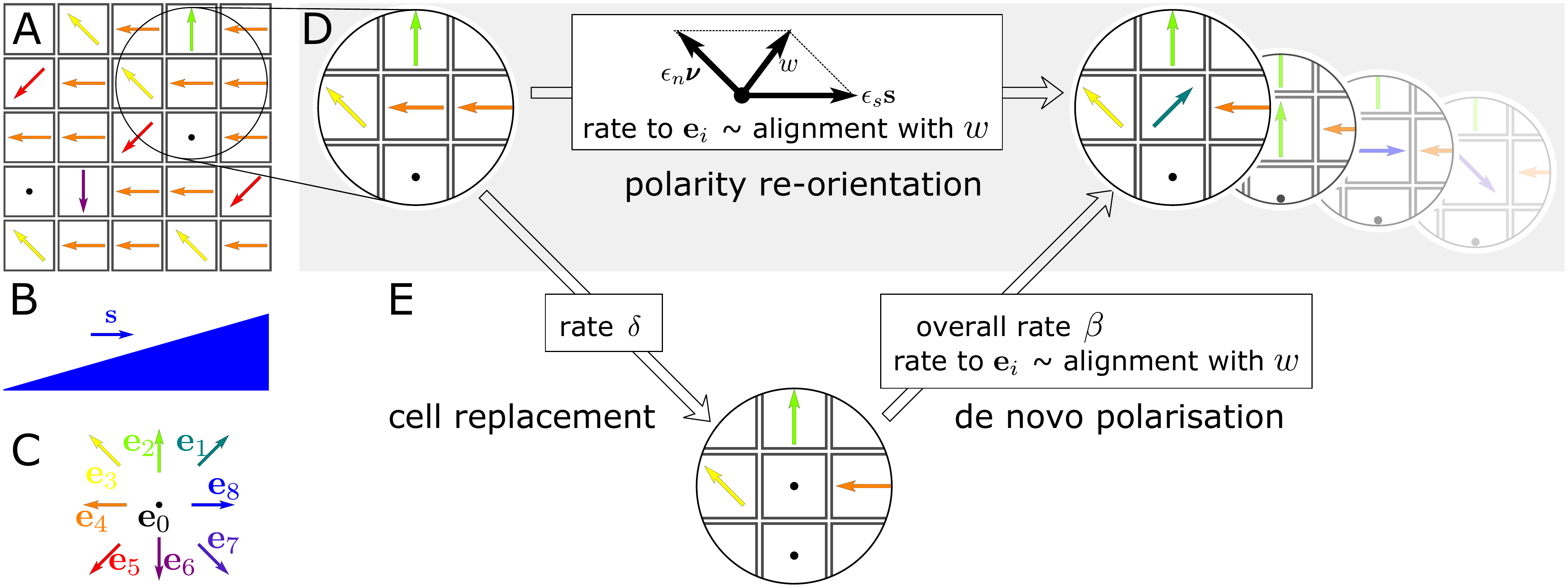}   
\caption{
The dynamically diluted alignment model, an Interacting Particle System (IPS) model of tissue polarity reorganisation. 
\textbf{A.} Each cell of a regular square lattice carries a polarity vector that initially preferentially points into a selected direction, here to the left. 
\textbf{B.} A uniform global signal~$\mathbf{s}$, e.g. the slope vector of a long range gradient (blue), opposes the initially dominant polarisation direction. 
\textbf{C.} Polarity directions are discretised into nine elementary states~$\mathbf{e}_0,\mathellipsis,\mathbf{e}_8$, shown with their color code.
\textbf{D,E.} State transitions are stochastic and affect one cell at a time, i.e. asynchronous update. 
\textbf{D.} Polarity reorientation tends to align cell polarity with the weighted ($\epsilon_n, \epsilon_s$) sum of the average polarisation vector~$\pmb{\nu}$ of the four direct neighbours and with the global signal~$\mathbf{s}$. Stacked copies on the right show potential outcomes of a single update with different chances. 
\textbf{E.} Additionally, dynamic dilution of tissue polarity results from cellular transitions to the null polarisation vector~$\mathbf{e}_0$ upon cell replacement at rate~$\delta$, equally for all polarised cells, followed by de-novo polarisation at rate~$\beta$. A cell selected for de-novo polarisation reaches a particular state $\mathbf{e}_1, \mathellipsis, \mathbf{e}_8$ with a chance distributed as for reorientation in \textbf{D}. %depending on the state's alignment with the reference orientation~$\mathbf{w}$, see~eq.~\eqref{eq:Def_rate_unpol_to_pol}. 
Model time is continuous and non-dimensionalised using $1 / \gamma$ as model unit time. 
}
\label{fig:ModelDef}
\end{figure}

%1 State space and configuration
Each cell is equipped with one of nine polarisation states, see fig.~\ref{fig:ModelDef}A, as follows. 
Thereby, the highly asymmetric concentration profile of PCP complexes along the cell membrane of a \emph{polarised} cell, that determines the cell's polarity orientation, is abstracted as one unit vector per cell pointing 
% from the cell center
towards the highest membrane accumulation of a selected PCP component
The directions of the unit vectors are discretised yielding the eight states 
\begin{align}  \label{eq:Def_e_i}
\mathbf{e}_i \define  \left(  \cos \left( i \pi / 4 \right), \sin \left( i \pi / 4 \right) \right), \qquad i =1,\mathellipsis 8 .
\end{align}
Na{\"\i}ve cells before complete polarity establishment are considered \emph{unpolarised} and are represented by a ninth state $\mathbf{e}_0 \define \left(0,0\right)$.
Thus, a cell at node $z \in S$ has polarisation state $\eta_z$ which is an element of $W \define \left\{ \mathbf{e}_i, \, i = 0, \mathellipsis, 8 \right\}$.
The state space of the whole system is $W^S$.
An element $\pmb{\eta} = \left( \pmb{\eta}_z \right)_{z \in S} \in W^S$ of the state space is called configuration and describes the global state of the system.
%2-9 dynamics
The model dynamics comprise two processes acting on individual cells, polarity alignment and cell turnover. 
Polarity alignment in turn is directed by two signals, local neighbours' polarities and a global orienting signal. 
%3 neighbour influence
As asymmetric protein complexes bridge adjacent cell membranes, polarisation of each cell tends to align with neighbours' polarisation vectors (see Introduction). 
In the model, a cell's neighbourhood is defined as those cells sharing a cell-cell~interface with that cell.
This is implemented by considering von Neumann neighbourhood in the square lattice which we complete with periodic boundaries. 
Assuming approximately equal lengths of cell-cell~interfaces, we use the equally weighted average polarisation vector 
\begin{align}  \label{eq:Def_nu}
\pmb{\nu}_z \left( \pmb{\eta} \right) \define \frac{1}{\# N_z} \sum_{x \in N_z} \pmb{\eta}_x , \qquad N_z \define \left\{ \textnormal{neighbours of } z \right\},  \qquad z \in S
\end{align} 
of neighbours $x$ to node $z$ as the local director of polarity alignment. 
Here $\# A$ denotes the number of elements of any set $A$.
%4 global gradient
Additionally, a global vector of polarity alignment is considered, representing the slope orientation $\mathbf{s} = \left(s_x, \, s_y \right)$ of a tissue-scale gradient.
%5 reference orientation
The local director $\pmb{\nu}_z \left( \pmb{\eta} \right)$ and the global director $\mathbf{s}$ can be differently weighted by a neighbour coupling strength $\epsilon_n \geq 0 $ and a coupling strength to the global signal $\epsilon_s \geq 0$, respectively. 
Both weighted vectors are then summed vectorially to yield the reference orientation 
\begin{align}  \label{eq:Def_w}
\mathbf{w}_z \define \epsilon_n \pmb{\nu}_z \left( \pmb{\eta} \right) + \epsilon_s \mathbf{s}
\end{align}
for node $z$, see fig.~\ref{fig:ModelDef}B.    
Considering a \emph{polarised} cell over time, a change of polarity to any new direction is modelled as more probable the more the new direction is aligned with the reference orientation~$\mathbf{w}$, but it is assumed to be \emph{independent} of the current polarisation direction of the considered cell. 
We deliberately consider abrupt changes in polarisation direction, because protein complexes bridging pairs of membranes from neighboring cells cannot shift across cell vertices but disassemble at a given cell interface and assemble anew at another interface.
The degree of alignment is measured by the standard scalar product $\left\langle \cdot, \cdot \right\rangle$ in $\mathbb{R}^2$. 
%6 definition of rates, part 1 (polarised-->polarised)
The rate $c_{z}$ for changing polarisation direction in node $z \in S$ from polarised state $\pmb{\eta}_z \in W \setminus \left\{ \mathbf{e}_0 \right\} $ to another polarised state $\mathbf{e}_i$, $i =1,\mathellipsis, 8 $, while keeping all other nodes unchanged is then defined as
\begin{align}     \label{eq:Def_rate_pol_to_pol}
\operatorname{c}_z  \left( \pmb{\eta} , \mathbf{e}_i \right)  &  \define \gamma \cdot  \exp \left\langle \mathbf{e}_i, \, \mathbf{w} \right\rangle   \notag  \\
{} & = \gamma \cdot \exp \left\langle \mathbf{e}_i, \, \epsilon_n \cdot \pmb{\nu}_z \left(\pmb{\eta} \right) + \epsilon_s \cdot \mathbf{s} \right\rangle \notag  \\
{} & = \gamma \cdot \exp \left(  \epsilon_n \left\langle \mathbf{e}_i, \, \pmb{\nu}_z \left(\pmb{\eta} \right)  \right\rangle + \epsilon_s \left\langle \mathbf{e}_i, \,  \mathbf{s} \right\rangle \right)
, \qquad \textnormal{for } \pmb{\eta}_z \neq \mathbf{e}_0, \, i=1,2,\mathellipsis,8
\end{align}
where parameter $\gamma$ gives the overall pace of polarity reorientation, in analogy to previous alignment models \cite{Vicsek1995,Peruani2011,Bussemaker1997}.

%7 cell turnover
Additionally, cell turnover (ageing and replacement by na{\"\i}ve cells) occurs independently of polarisation direction. 
%8 definition of rates, part 2 (loss of polarisation)
In the model, we let a polarised cell~$z \in S$ with $\pmb{\eta}_z \in W \setminus \left\{ \mathbf{e}_0 \right\} $ change into the unpolarised state $\mathbf{e}_0$ with death rate~$\delta \geq 0$, 
\begin{align} \label{eq:Def_rate_pol_to_unpol}
\operatorname{c}_{ z }  \left( \pmb{\eta} , \mathbf{e}_0 \right)  \define  \delta   ,   \qquad \textnormal{for } \pmb{\eta}_z \neq \mathbf{e}_0 .  
\end{align}
%9 definition of rates, part 3 (de-novo polarisation)
The establishment of any polarisation direction from scratch in an unpolarised cell is modelled with de-novo polarisation rate $\beta \geq 0$. 
By setting
\begin{align} \label{eq:Def_rate_unpol_to_pol}    
\operatorname{c}_z  \left( \pmb{\eta} ,  \mathbf{e}_i \right)  \define  \beta \cdot \frac{ \gamma \cdot \exp \left(  \left\langle \mathbf{e}_i, \, \epsilon_n \cdot \pmb{\nu}_z \left(\pmb{\eta} \right) + \epsilon_s \cdot \mathbf{s} \right\rangle \right) }{ \sum_{k=1}^{8} \gamma \cdot \exp \left(  \left\langle \mathbf{e}_k, \, \epsilon_n \cdot \pmb{\nu}_z \left(\pmb{\eta} \right) + \epsilon_s \cdot \mathbf{s} \right\rangle \right) }       ,   \qquad \textnormal{for } \pmb{\eta}_z  = \mathbf{e}_0 , \, i =1,\mathellipsis, 8 \, ,
\end{align}
the polarisation directions after de-novo polarisation are distributed as those in a re-alignment step, cf.~eq.~\eqref{eq:Def_rate_pol_to_pol} and note that $\gamma$ cancels. 
Taken together, the model \cref{eq:Def_e_i,eq:Def_nu,eq:Def_w,eq:Def_rate_pol_to_pol,eq:Def_rate_pol_to_unpol,eq:Def_rate_unpol_to_pol} define the transition rates of a continuous time Markov chain $\left(\pmb{\eta}  \left( t \right) \right)_{t \geq 0}$ or more specifically an IPS \cites{Liggett1985}{IPS_in_EncyclopediaSystemsBiology}{KlaussVoss-Boehme2005}{KlaussVoss-Boehme2008}, which we call the \textit{dynamically diluted alignment model}. 
See fig.~\ref{fig:ModelDef} for an illustration of the model dynamics. 
%10 repeat question in mathematical framework
This dynamically diluted alignment model allows to study the effects of cell turnover on polarity patterns.
Specifically, we ask whether and how polarity patterns with coherent initial polarisation counter-directional to the global signal~$\mathbf{s}$ reorganise, to resolve the conflict between local and global directional cues, and what the time requirement is if they do so.

\subsection{Parametrisation}  \label{subsec:Parametrisation}
%1 remind of model parameters
The model parameters are $\delta$, $\beta$, $\gamma$, $\epsilon_n$, $\epsilon_s$, $\mathbf{s}$, see sec.~\ref{subsec:Model_Def} and fig.~\ref{fig:ModelDef}, and the end time $t_{max}$ of a simulation.
The complete list of symbols is given in suppl.~table~\ref{tab:ListOfSymbols}.
%2 Discuss time scale
We dedimensionalise by choosing $1 / \gamma$ as time unit of the model.
Hence the rates $\beta$ and $\delta$ become dimension-less parameters, and we omit $\gamma$ whenever possible.
%We non-dimensionalise the time scale by setting $t_{max} = 1 \left[ T \right]$, such that the final observation time~$T$ of a comparable \textit{in vivo} experiment corresponds to the time unit in the model. 
%Then any rate directly translates into the expected number of events during such an experiment.
The cell death rate~$\delta$ and de-novo polarisation rate~$\beta$ may not be directly experimentally accessible. 
%3 equations that link model parameters (\beta, \delta) to experimental observables (\tau, p_{eq})
However, there are two quantities that are measurable, 
the average duration~$\tau$ of a cell's complete cycle through unpolarised and polarised states and the fraction of polarised cells, which can together be used to determine $\delta$ and $\beta$ as follows. 
First, since a single polarised cell looses polarisation at rate $\delta$, it remains polarised on average for $1/\delta$ time units. 
Analogously, unpolarised cells remain so for $1/ \beta$ time units on average.
Hence the average duration~$\tau$ of a complete cycle through unpolarised and polarised states per cell is related to $\beta$ and $\delta$ by
\begin{align}  \label{eq:tau}
\frac{ \tau}{ \gamma} = \frac{1}{\beta} + \frac{1}{\delta} 
\end{align} 
where the denominator $ \gamma$ accounts for the unit model time.

Second, the fraction of polarised cells~$p_{eq}$ among all cells equals, after any initial transients have decayed, the ratio of the expected duration of the polarised state over the expected duration of the whole cycle of a cell, 
\begin{align}  \label{eq:frac_pol}
p_{eq} = \frac{1/\delta}{\tau / \gamma} = \frac{\beta}{\beta + \delta}        .
\end{align}
Solving \cref{eq:tau,eq:frac_pol} for $\left(\beta, \, \delta \right)$ yields
\begin{align}    \label{eq:beta_and_delta}   
\beta & = \frac{\gamma}{\tau} \cdot \frac{1}{1 - p_{eq}} , & { }  \delta & =  \frac{\gamma}{\tau} \cdot  \frac{1}{p_{eq}}   .
\end{align}

%4 pick values of (\beta, \delta) that seem physiological
We explore the model behaviour for $\beta \in \left\{0.1,1,10\right\}$.
For each value of $\beta$ we vary $\delta$ between $0$ and $10 \beta$.
By eq.~\eqref{eq:frac_pol}, this corresponds to $p_{eq}$ ranging from $1$, where all cells are polarised (the case of dynamic Potts model
without dilution), to a fraction of polarised cells around $0.099$, effectively covering the fraction of polarised cells in all known polarised epithelia. % So we surely cover the true value of $p_{eq}$.
The resulting values of $\tau$ range in $\left[ 0.11, \, \infty \right]$. %Note that \delta=0 implies infinite lifespan.
For planarians, an experimental observation time $T = \SI{14}{\day}$ \cite{Rink2013} then corresponds to a cycle length from immigration into the tissue via polarisation until the cell's death ranging from $ \SI{1.5}{\day}$ upwards, which is realistic. 
%5 Further parameter fixation
Below, the impact of the dimensionless weights $\epsilon_n$, $\epsilon_s$ that represent cellular sensitivities to neighbours' polarity and global signal, respectively, will be studied in detail. 
Since the reference orientation~$\mathbf{w} \define \epsilon_n \pmb{\nu} \left( \pmb{\eta} \right) + \epsilon_s \mathbf{s}$ is a weighted sum of the two directional cues, it suffices to keep $\epsilon_s = 1$ fixed and vary $\epsilon_n \in \left[ 0, 5 \right]$. 
The model is symmetric w.r.t. %with respect to
the discretised directions $\mathbf{e}_1, \mathellipsis, \mathbf{e}_8$, so the $\mathbf{e}_8$ direction can be chosen to coincide with the direction of the global signal, i.e. $\mathbf{s} = \left(1, \, 0 \right) = \mathbf{e}_8$.

%6 initial configuration
The initial configuration shall represent a homeostatic tissue with fraction~$p_{eq}$ of polarised cells, where the dominant polarisation direction is opposite to the global signal~$\mathbf{s}$.
Therefore, we assign the initial state $\mathbf{e}_0$ with probability $1 - p_{eq} = \delta / \left( \beta + \delta \right)$ to each cell independently
and set the initial states of polarised cells as if each cell had experienced prior to simulation start coherent global and local directors $\mathbf{s}_{init} = \pmb{\nu}_{init} = \left(-1, \, 0 \right) =  \mathbf{e}_4 = - \mathbf{s}$, opposite to $\mathbf{s}$.
The latter means that each polarised cell is assigned $\mathbf{e}_k$ with probability 
\begin{align}  \label{eq:initial_distribution}
\frac{ \beta }{ \left( \beta + \delta \right) } \frac{ 1 }{ Z } \exp \left\{ \left(\epsilon_n +  \epsilon_s \right) \left\langle \mathbf{e}_k, \,  \mathbf{e}_4  \right\rangle  \right\}  , \qquad  k=1, \mathellipsis , 8 ,
\end{align}
independently, where $ Z \define \sum_{i=1}^{8}  \exp \left\{ \left(\epsilon_n +  \epsilon_s \right) \left\langle \mathbf{e}_i, \,  \mathbf{e}_4  \right\rangle  \right\}$ is a normalisation constant.
This way, cells are initially coherently polarised with main polarisation direction $\mathbf{e}_4 =  \left(-1, \, 0 \right)$, which indeed conflicts with $\mathbf{s} = \left(1, \, 0 \right)$, see fig.~\ref{fig:Results_IPS_size100}A.

\subsection{Observables}  \label{subsec:Observables}
%1 Define observable: mean polarisation
To decide whether the polarity pattern adapts to the global signal, and if so to quantify the time requirement for reorientation, we track the mean polarisation 
\begin{align} \label{eq:Def_orderParam}
\mathbf{p} \left( \pmb{\eta} \left( t \right) \right) \define \frac{1}{\# S  } \sum_{z \in S } \pmb{\eta}_z \left(t\right)  \quad \in \mathbb{R}^2  
\end{align}
of an evolving configuration $\pmb{\eta} \left( t \right)$ during the simulation.
The behaviour of this polar alignment order parameter is best described in terms of modulus $\left\| \mathbf{p} \right\| $ and angle to the positive $x$-axis $\operatorname{ang} \left( \mathbf{p} \right)$.
Due to the stochasticity of the model, both modulus and angle fluctuate, but these variations decrease with increasing lattice size, see SI~figure~\ref{fig:suppl:Results_IPS_size20}.
%2 discuss behaviour of mean polarisation: modulus
The modulus $\left\| \mathbf{p} \right\| \in \left[0,1\right]$ characterises the degree of alignment among all cells' polarisation directions. 
If $\left\| \mathbf{p} \right\| = 0$ then no globally dominant polarisation direction exists, whereas in case of $\left\| \mathbf{p} \right\| = 1$ all cells are polarised into the same direction. 
However, $\left\| \mathbf{p} \right\| = 1$ requires that all cells are polarised. 
Due to cell turnover, the actual fraction of polarised cells 
\begin{align}    \label{eq:Def_p_p}
p_p \define \frac{\# \left\{ \textnormal{polarised cells} \right\} }{\# S}  = 1 -  \frac{\# \left\{ z \in S; \, \eta_z = \mathbf{e}_0 \right\} }{\# S}
\end{align}
fluctuates around $p_{eq}$ and obeys $\left\| \mathbf{p} \right\| \leq p_p \leq 1$.
%2b discuss alternative measures
Hence $\left\| \mathbf{p} \right\|  / p_p \leq 1$ characterises the degree of alignment among the \emph{polarised} cells where equality holds if and only if all polarised cells share one direction. 
%3 discuss behaviour of mean polarisation:  angle
The angle $\operatorname{ang} \left( \mathbf{p} \right)$ indicates the predominant polarisation direction, and exhibits switching behaviour if the polarity pattern reorients.
Hence a high value of $\left\| \mathbf{p} \right\| $ together with an angle $\operatorname{ang} \left( \mathbf{p} \right)$ approximately oriented parallel to the global vectorial signal~$\mathbf{s}$ will inform us that polarity reorientation has occurred.
Then local and global signals are coherent and the configurations are in a stochastic dynamic equilibrium.

%4 Introduce \tmo
To quantify the time requirement of reorientation, we monitor the time needed to reach the dynamic equilibrium. 
We observe that, as a prerequisite for polarity pattern reorientation, the degree of alignment measured by $\left\| \mathbf{p} \right\| $ diminishes until $\left\| \mathbf{p} \right\| $ attains a distinct minimum, and that $\operatorname{ang} \left( \mathbf{p} \right)$ undergoes the fastest change around that time, cf.~fig.~\ref{fig:Results_IPS_size100}B. 
In addition, the first phase in which the initially coherent polarity pattern resolves into a minimally ordered transient state, is of particular interest to study the influence of conflicting signals, 
while the subsequent evolution of a disordered system towards an ordered state due to the global signal has been studied before \cite{BurakShraiman2009}. 
Therefore, we use the time of minimal order
\begin{align} \label{eq:Def_t_mo}
\tmo \define \operatorname {argmin}_{t \in \left[0, 1 \right]} \left\| \mathbf{p} \left( t \right) \right\|  
\end{align}
as the characteristic, statistically robust time for conflict resolution in tissue polarity reorganisation, cf.~fig.~\ref{fig:Results_IPS_size100}B,C.

\subsection{Model analysis}

\subsubsection{Simulation}\label{subsec:Simulation}
To sample from the trajectories of our stochastic model we employ the exact stochastic simulation algorithm by Gillespie \cite{Gillespie1977} in an efficient implementation for IPS \cites{KlaussVoss-Boehme2005}{KlaussVoss-Boehme2008}. 
Three of six parameters are held fixed as described in section~\ref{subsec:Parametrisation}: $\gamma=1$ by dedimensionalisation, $\mathbf{s} = \left( 1,0 \right)$, $\epsilon_s = 1$.
The initial configuration at $t=0$ is specified by eq.~\eqref{eq:initial_distribution}.
The other three parameters are varied as $\epsilon_n = 0, 0.5, 1.0, \mathellipsis, 5.0$, $\beta = 0.1, \, 1, \, 10$, $\delta=0, 0.2 \beta, 0.4 \beta, \mathellipsis, 10.0 \beta$.
Simulations are carried out on a $100 \times 100$~lattice with periodic boundary conditions until simulated time exceeds $t_{max} = 1$.
See fig.~\ref{fig:Results_IPS_size100}A and suppl.~movie for an example simulation, and suppl.~fig.~\ref{fig:suppl:Results_IPS_size20} for a justification that the lattice size is sufficient. 

\subsubsection{Mean-field analysis} \label{subsubsec:mean-fieldAnalysis}
%1 objective of the subsection
Using mean-field approximation, we derive an ODE which approximates the temporal evolution of $\mathbf{p} \left( t \right) = \mathbf{p} \left( \pmb{\eta} \left( t \right) \right)$.
See supplement~\ref{suppl:mean-fieldAnalysis} for more details of the following derivations.
%2 define fractions as observable in the IPS
Denote the fractions $\mathbf{a} (t) = \left( a_0 (t), \mathellipsis , a_8 (t) \right)$ of nodes in states $\mathbf{e}_0, \mathellipsis , \mathbf{e}_8$ at time $t$ in the IPS model by 
\begin{align}  \label{eq:Def_a_i}
a_i \left( t \right)  \define  a_i \left( \pmb{\eta} \left( t \right) \right)  \define \frac{ \# \left\{ z \in S, \, \pmb{\eta}_z \left( t \right) = \mathbf{e}_i \right\}  }{\# S }  , \qquad i=0, \mathellipsis , 8 \, .
\end{align}
Then the mean polarisation vector can be expressed as  
\begin{align}  \label{eq:representations_of_p}
\mathbf{p} \left( t \right) = \left( p_x \left( t \right), \, p_y \left( t \right) \right)^{\transpose} =  \sum_{i=1}^{8} a_i \left( t \right) \cdot \mathbf{e}_i  
\end{align}
and the fraction of polarised cells is $p_p \left( t \right) = 1- a_0 \left( t \right) = \sum_{i=1}^{8} a_i \left( t \right)$. 

%3 Mean-field approximation, rates from IPS become rates $r_k$
\noindent The mean-field assumption (MFA) simplifies the rates $\operatorname{c}_z  \left( \pmb{\eta} , \mathbf{e}_i \right)$ defined in eqs.~\eqref{eq:Def_rate_pol_to_pol} and \eqref{eq:Def_rate_unpol_to_pol} to (cf.~supplement~\ref{suppl:mean-fieldAnalysis})
\begin{align}  \label{eq:MFA_rate_pol_to_pol_AND_unpol_to_pol}  %merges {eq:MFA_rate_pol_to_pol} and {eq:MFA_rate_unpol_to_pol}    
\operatorname{c}_z  \left( \pmb{\eta} , \mathbf{e}_i \right)    \stackrel{MFA}{\approx}     \begin{cases}  r_i \left( \mathbf{a} \right) = \exp \left( \left\langle \mathbf{e}_i, \,  \epsilon_n \cdot M \mathbf{a} + \epsilon_s \cdot \mathbf{s} \right\rangle \right),   \qquad &  \textnormal{if } \pmb{\eta}_z \neq \mathbf{e}_0, \, i=1,\mathellipsis,8   \\       \beta \frac{r_i \left( \mathbf{a} \right)} {\sum_{k = 1}^{8}  r_k \left( \mathbf{a} \right) }   =    \beta \frac{r_i \left( \mathbf{a} \right)} {R \left( \mathbf{a} \right) }      , \qquad &  \textnormal{if } \pmb{\eta}_z  = \mathbf{e}_0 , \, i =1,\mathellipsis, 8   .    \end{cases} 
\end{align}
%The approximations herein are exact for $\epsilon_n = 0$, and the approximation error increases with $\epsilon_n$.

In the limit for increasing lattice size the $a_i$'s become continuous quantities and their dynamic behaviour can be described by an ODE system \cite{vanKampen1997,Boettger2015,Hohmann2013}
%4 ODE for (a_0, a_1, ... , a_8)
\begin{align} \label{eq:ODE_expressed_in_rR_a_0still_free} 
\left.    \begin{aligned}  
    \frac{\mathrm{d} \MFA{a}_i }{\mathrm{d} t }   &  =   - \MFA{a}_i  \cdot   \left( \delta + \sum_{ k =1 }^{8} r_k \left( \MFA{ \mathbf{a} } \right)  \right)    +    \MFA{a}_0 \cdot \beta \frac{r_i \left(  \MFA{ \mathbf{a} } \right)} {\sum_{k = 1}^{8}  r_k \left(  \MFA{ \mathbf{a} } \right)}  +   \sum_{ k =1}^{8}  \MFA{a}_k \cdot r_i \left(  \MFA{ \mathbf{a} } \right)      \\   
    { }   &  =   - \MFA{a}_i  \cdot   \left( \delta + R \left(  \MFA{ \mathbf{a} } \right)  \right) +  \left[ \MFA{a}_0 \cdot \frac{\beta}{R \left(  \MFA{ \mathbf{a} } \right)} + \left(1-\MFA{a}_0\right)   \right] r_i \left(  \MFA{ \mathbf{a} } \right)   , \qquad i=1,\mathellipsis, 8      \\
    \frac{\mathrm{d} \MFA{a}_0 }{\mathrm{d} t }  &   = - \MFA{a}_0 \cdot \beta + \delta \cdot \left(1 - \MFA{a}_0 \right)  =   \delta  - \left(\beta + \delta  \right) \MFA{a}_0      .     
\end{aligned}     \right\}    
\end{align}
Here $\MFA{ \mathbf{a} }$ and $\MFA{a}_i$ denote the counterparts of $ \mathbf{a}$ and $a_i$ under mean-field approximation (MFA).
We call eq.~\eqref{eq:ODE_expressed_in_rR_a_0still_free} the mean-field model and note that the overall error of approximation introduced in eq.~\eqref{eq:MFA_rate_pol_to_pol_AND_unpol_to_pol} increases with $\epsilon_n$, $\epsilon_s$, and the fraction of polarised cells~$p_p$.
The fraction of unpolarised cells $\MFA{a}_0 \left( t \right)$ tends to the unique, globally attracting equilibrium $\MFA{a}_0^{\ast} = \frac{\delta}{\beta + \delta} $, 
which is in perfect agreement with the dynamic equilibrium $p_{eq} = \frac{\beta}{\beta + \delta} = 1 - \MFA{a}_0^{\ast}$ of death and de-novo polarisation in the original IPS (eq.~\eqref{eq:frac_pol}). 
%5 ODE for (a_1, ... , a_8) with a_0 fixed
This equilibrium $\MFA{a}_0 = \MFA{a}_0^{\ast} = \frac{\delta}{\beta + \delta} $ simplifies the ODE system~\eqref{eq:ODE_expressed_in_rR_a_0still_free} to
\begin{align}  \label{eq:ODE_expressed_in_rR}
\frac{\mathrm{d} \MFA{a}_i }{\mathrm{d} t }  &  =   \left( \delta + R \left( \MFA{ \mathbf{a} } \right) \right) \cdot \left(  -\MFA{a}_i  +   \frac{\beta}{\beta + \delta}  \frac{ r_i \left( \MFA{ \mathbf{a} } \right) }{ R \left( \MFA{ \mathbf{a} } \right) }   \right)  , \qquad i=1,\mathellipsis, 8  ,  
\end{align}
which can be summed to an ODE for $ \MFA{ \mathbf{p} }$, see suppl.~eq.~\eqref{suppl:ODE_for_p_expressed_in_rR}.
Solutions of \eqref{eq:ODE_expressed_in_rR} preserve the symmetry of the initial condition with respect to the \mbox{$x$-axis}, that was imposed by setting $\mathbf{s}=\left(1,0\right)$, $a_1 \left(0\right) = a_7 \left(0\right)$, $a_2 \left(0\right) = a_6 \left(0\right)$ and $a_3 \left(0\right) = a_5 \left(0\right)$ (cf.~eq.~\eqref{eq:initial_distribution}).
For such a symmetric initial condition, it holds $\MFA{ \mathbf{p} }  \left( t \right) = \left( \MFA{ p }_x \left( t \right), 0 \right) $ for all times $t >0$.
Numerical solutions of eq.~\eqref{eq:ODE_expressed_in_rR} are shown in section~\ref{subsec:NumericalSolutionOfMean-fieldODE}.

\subsubsection{Linearisation of mean-field model}  \label{subsubsec:ODELinearisation}
%0 Motivation
One can approximate the non-linear mean-field model \eqref{eq:ODE_expressed_in_rR} further, see supplement~\ref{suppl:ODELinearisation}, to obtain an analytically tractable ODE
\begin{align}   \label{eq:linearisedODE}
\frac{\mathrm{d} \linMFA{a}_i }{\mathrm{d} t }   &  =   \left( \delta + 8 \right)  \left(    - \linMFA{a}_i  +  \frac{1}{8} \frac{\beta}{\beta + \delta}  \left[  1 +  \epsilon_n  \sum_{k=1}^{8}   \linMFA{a}_k \left\langle \mathbf{e}_k, \, \mathbf{e}_i \right\rangle + \epsilon_s \left\langle \mathbf{s}, \, \mathbf{e}_i \right\rangle   \right]  \right) , \qquad i=1,\mathellipsis, 8     
\end{align} 
where an overline over symbols indicates this second approximation.
Again, an ODE for $ \linMFA{ \mathbf{p} }$ (suppl.~eq.~\eqref{suppl:linearised_ODE_for_p_x_p_y_decoupled_linMFa_0_steady}) follows.
The analytical solution (suppl.~eq.~\eqref{suppl:t_mo_linearised_MF_most_general}) implies that
polarity reverses in the linearised mean-field model only for $\epsilon_n \frac{\beta}{\beta + \delta} < 2$.
In this case, the time of minimal order is uniquely determined as (see suppl.~eq.~\eqref{suppl:t_mo_linearised_MF_general})
\begin{align}  \label{eq:t_mo_linearised_MF_general}   % assuming $s_x = 1 $
\tmolinMF & =  \frac{2}{ \left( \delta + 8 \right) \left( 2 - \epsilon_n \frac{\beta}{\beta + \delta} \right)}   \ln \left( 1 +   \frac{2 + \epsilon_n \frac{\beta}{\beta + \delta}}{\epsilon_s}
          \cdot   \frac{ - \linMFA{ p }_x  \left( 0 \right)}{ s_x \frac{\beta}{\beta + \delta}}  \right)     .
\end{align}

%=========================================================================================================================================
\section{Results} \label{sec:Results}

\subsection{Simulation of the dynamically diluted alignment model}  \label{subsec:Results:IPSdata}

%The dynamically diluted alignment model as defined in section~\ref{subsec:Model_Def} and Fig.~\ref{fig:ModelDef} was simulated as described in section~\ref{subsec:Simulation}.

\paragraph{Time course in simulations.}
%0 Reorientation happens for all parameter values tested
For all parameter combinations studied, the polarity pattern resolved the initial conflict within the simulated time window $t \in \left[0, t_{max} \right] = \left[0, 1\right]$ by reversing the main polarisation direction and adapting to the global signal. 
Even for the largest chosen neighbour coupling strengths $\epsilon_n = 5.0$, we never observed the frustrated initial condition to persist.
The time courses of $ \left\| \mathbf{p} \right\|$ and $\operatorname{ang} \left( \mathbf{p} \right)$ exhibit several common characteristics independent of the specific parameter sets, described as follows.

\begin{figure}
{
  \includegraphics[width=1.0 \textwidth, keepaspectratio]{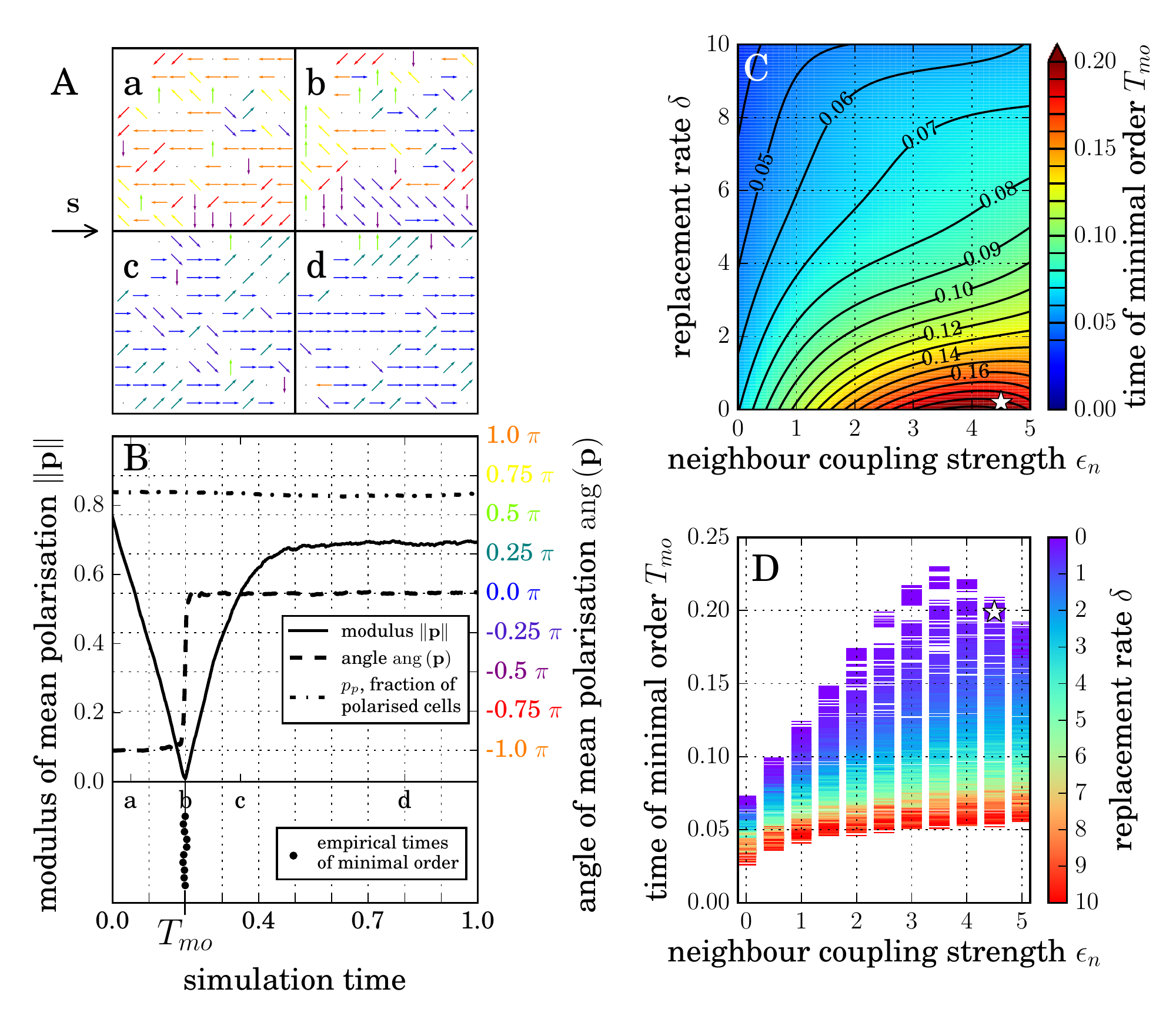} 
} 
\caption{
The time of minimal order characterises the plasticity of tissue polarity patterns. 
\textbf{A.} Snapshots of a typical simulation of the IPS model at times~0.05,~0.2,~0.5,~0.8 (\textbf{a-d}), see fig.~\ref{fig:ModelDef}C for color code and suppl.~movie. 
Zoomed details of a $100 \times 100$~lattice with periodic boundaries, $\epsilon_n=4.5$, $\delta=0.2$, $\beta=1$, $\epsilon_s=1$, $\mathbf{s}=\left(1,0\right)$. 
\textbf{B.} Mean polarisation vector~$\mathbf{p}$ depicted as modulus~$\left\| \mathbf{p} \right\| $ (solid, left axis) and angle~$\operatorname{ang} \left( \mathbf{p} \right)$ (dashed, right axis, color code as in A and fig~\ref{fig:ModelDef}C). 
The distinctive minimum of $\left\| \mathbf{p} \left( t \right) \right\|$ defines the time of minimal order~$T_{mo,i}$ %$\tmo_{,i}$
of time course $i \in \mathbb{N}$. 
The fraction of polarised cells $p_p$ (dash-dotted, left axis) fluctuates around $p_{eq} = 0.8{\bar{3}}$.
\textbf{C,D.} Simulation results for mean $\tmo$ of 25 repetitions shown as heatmap with contourlines (isotemporales at marked levels, \textbf{C}) and all data points (\textbf{D}), fixed parameters as in A,B. Asterisks denote parameter values of panels A,B. Note inverted color bar for $\delta$ in \textbf{D}.
}
\label{fig:Results_IPS_size100}  
\end{figure}

%1 description of exemplary simulation (fig 3A) 
Fig.~\ref{fig:Results_IPS_size100}A and suppl.~movie show an exemplary trajectory of the dynamically diluted alignment model together with the time course of the observable~$\mathbf{p}$ derived from eq.~\eqref{eq:Def_orderParam} in fig.~\ref{fig:Results_IPS_size100}B.
According to the initialisation (cf.~eq.~\eqref{eq:initial_distribution}), the system starts globally ordered where the dominant orientation~$\mathbf{e}_4 = \left(-1, 0 \right)$ is opposed to the global signal~$\mathbf{s}=\left( 1, 0 \right)$.
Single cells or small cell patches deviate from their initial polarisation direction (fig.~\ref{fig:Results_IPS_size100}Aa) 
and others follow until there is hardly any predominant polarisation direction around the time of minimal order (fig.~\ref{fig:Results_IPS_size100}Ab). 
This happens independent of lattice size, see suppl.~fig.~\ref{fig:suppl:Results_IPS_size20}.
%Cells adopting states $\mathbf{e}_1$, $\mathbf{e}_7$ and finally $\mathbf{e}_8$ change the dominant polarisation direction from $\mathbf{e}_4$ to $\mathbf{e}_8$.
In succession, the polarity pattern approaches a state of well aligned polarisation directions, dominated by $\mathbf{e}_8$ in coherence with the global signal (fig.~\ref{fig:Results_IPS_size100}Ac).
These configurations form a dynamic equilibrium that persists (cf. fig.~\ref{fig:Results_IPS_size100}Ac, \ref{fig:Results_IPS_size100}Ad).
The transition from alignment among cells conflicting with the global signal to alignment with the global signal happens via a disordered state when each polarisation direction is approximately equally abundant, see suppl.~fig.~\ref{fig:suppl:Results_IPS_size100_beta1_statistics}A.

%2 Description of observable during simulation (fig 3B) 
These dynamics are well recapitulated in the time course of the polarity alignment order parameter~$\mathbf{p}$, see fig.~\ref{fig:Results_IPS_size100}B and section~\ref{subsec:Observables}.
Its modulus $\left\| \mathbf{p} \right\| $ starts at a high value not above $ \frac{\beta}{\beta + \delta}$ and decreases to a distinct minimum at time of minimal order~$\tmo$ as more and more cells leave their initial directions.
The modulus $\left\| \mathbf{p} \right\| $ increases again immediately after attaining the minimum as a growing majority of cells adopts states $\mathbf{e}_1$, $\mathbf{e}_7$ and finally $\mathbf{e}_8$ when $\left\| \mathbf{p} \right\| $ reaches a plateau.
The angle $\operatorname{ang} \left( \mathbf{p} \right)$ remains almost constant~$\approx - \pi$ up to shortly before $\tmo$, increases steeply around $\tmo$ to a level of~$\approx 0$ around which it then fluctuates.
Note that the switching in $\operatorname{ang} \left( \mathbf{p} \right)$ is also possible as a decrease from~$\approx \pi$ to~$\approx 0$ for symmetry reasons, cf. suppl.~fig.~\ref{fig:suppl:Results_IPS_size20}.
Hence the time of minimal order~$\tmo$ is distinguished not only by the least degree of alignment $\left\| \mathbf{p} \right\| $ in the polarity pattern, but also by the fast, switch-like change of the angle $\operatorname{ang} \left( \mathbf{p} \right)$.
This indicates that pattern reorientation takes place during the short phase of low polarity alignment around~$\tmo$, and that the previous decrease in order is a precondition.
Therefore $\tmo$ is appropriate to study the time requirement of polarity pattern reorientation.
%3 Statistics of \tmo are good
Moreover, the time of minimal order $\tmo$ is statistically robust.
It differs only slightly between sampled trajectories of $\mathbf{p} \left( t \right)$ from a fixed parameter set, see fig.~\ref{fig:Results_IPS_size100}B, lower panel, and suppl.~fig.~\ref{fig:suppl:Results_IPS_size100_beta1_statistics}B.

%4 Parameter dependence of \tmo
\paragraph{Parameter dependence of time of minimal order $\tmo$.}
As described in section~\ref{subsec:Parametrisation}, it suffices to vary replacement rate $\delta$ and neighbour coupling strength $\epsilon_n$ to explore the model behaviour while keeping de-novo polarisation rate $\beta$ and coupling strength to global signal $\epsilon_s$ fixed.
Fig.~\ref{fig:Results_IPS_size100}C,D show the parameter dependence of the time of minimal order~$\tmo =  \tmo \left( \delta, \epsilon_n \right)$ for $\beta=1$ and $\epsilon_s=1$. 
See suppl.~fig.~\ref{fig:suppl:Results_IPS_size100_beta0_1and1and10}A-C for $\beta=0.1$, $\beta=10$ and empirical standard deviations, and suppl.~fig.~\ref{fig:suppl:Results_IPS_size20} for lattice size $20 \times 20$.
First, we find that the time of minimal order grows with $\epsilon_n$. 
This is plausible as neighbour coupling hinders cells from breaking free from their initially frustrated polarisation direction.
Second, we observe a decline in the time of minimal order with growing replacement rate $\delta$.
This is plausible as well, since higher values of $\delta$ for fixed $\beta$ imply an increased fraction of unpolarised cells, cf. eq.~\eqref{eq:frac_pol}, and the average neighbour polarisation shrinks in absolute value. 
Therefore the reference orientation $\mathbf{w}$, which is the vectorial sum of global and local directors (see eq.~\eqref{eq:Def_w} and fig.~\ref{fig:ModelDef}B) shifts towards the global signal, which reduces the decelerating effect of neighbour coupling and accelerates reorientation.
For $\delta >> \beta$, $\tmo$ depends less strongly on $\epsilon_n$ as evident from the smaller slope of the corresponding red data points compared to all others in fig.~\ref{fig:Results_IPS_size100}D. 
Actually $\tmo$ might approach some plateau for $\delta \to \infty$ while $\beta = const$  since polarised cells become virtually isolated, cf.~eq.~\eqref{eq:frac_pol}. 
The isotemporales in fig.~\ref{fig:Results_IPS_size100}D connecting parameter sets with equal $\tmo$ highlight the decelerating effect of increased neighbour coupling and the acceleration by cell turnover.

\subsection{Numerical solution of the mean-field model}  \label{subsec:NumericalSolutionOfMean-fieldODE}
\begin{figure} 
{\includegraphics[width=1.0 \textwidth, height= \textheight, keepaspectratio]{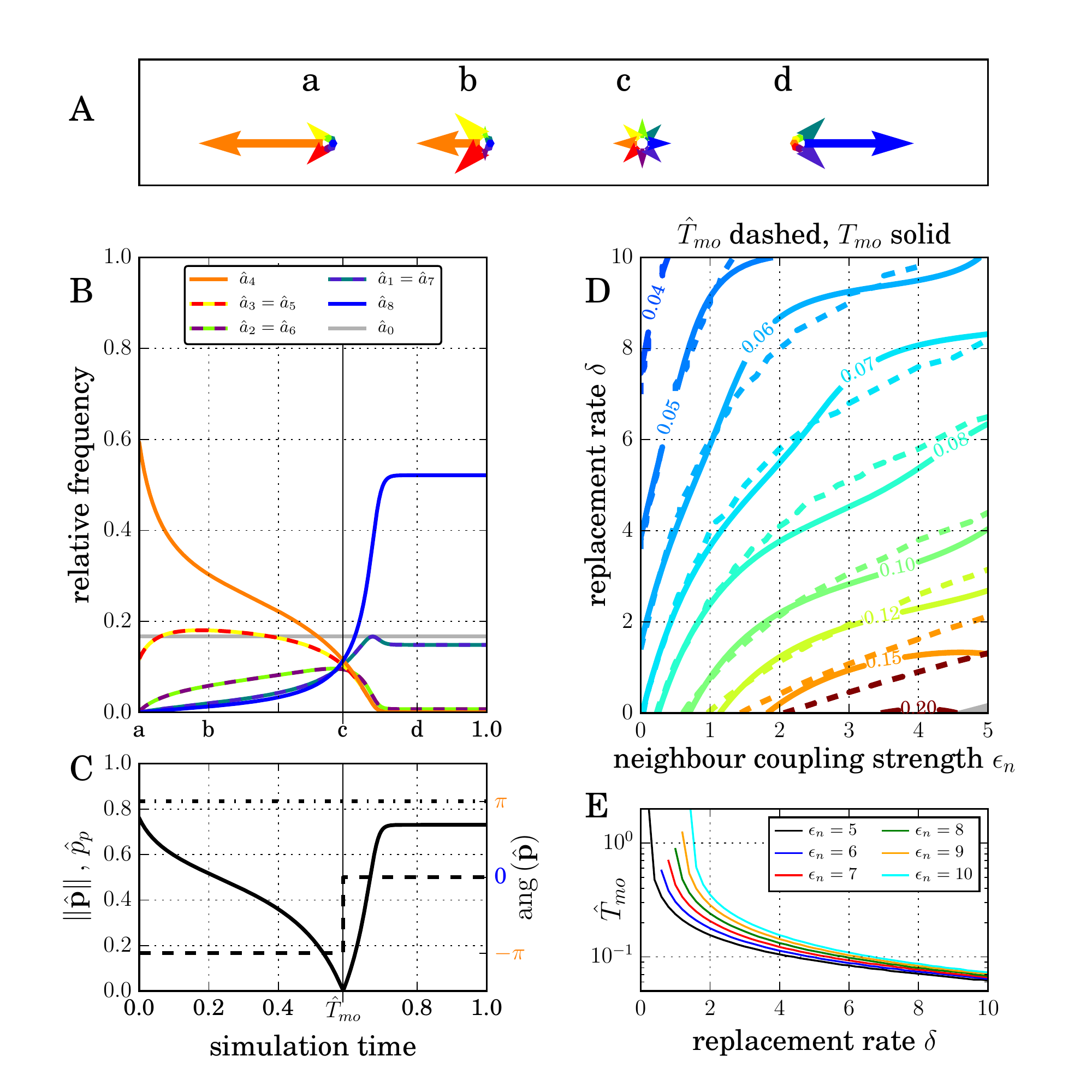}}
\caption{
Mean-field model. 
\textbf{A-C.} Numerical solution of \eqref{eq:ODE_expressed_in_rR} visualised as $\MFA{a}_i (t) \cdot \mathbf{e}_i$ in A, at times 0.0, 0.2, $\tmoMF$, 0.8 (\textbf{a-d}) with color code from fig.~\ref{fig:ModelDef}C; and as time course in B.
Parameter values are $\epsilon_s=1$, $\beta = 1$, $\delta = 0.2$ and $\epsilon_n=4.5$ as in fig.~\ref{fig:Results_IPS_size100}.
Panel C shows mean polarisation vector $\MFA{ \mathbf{p} } \left( t \right)$ with its modulus $\left\| \MFA{ \mathbf{p} } \right\|$ (solid, left axis) starting high, reaching a unique minimum with $\left\| \MFA{ \mathbf{p} } \right\| = 0$ that defines $\tmoMF$, and increasing again to a high plateau.
The angle $\operatorname{ang} \left( \MFA{ \mathbf{p} } \right)$ (dashed, right axis) switches from $-\pi$ to $0$ at $\tmoMF$. 
The fraction of polarised cells (dash-dotted) remains constantly $\MFA{p}_p = p_{eq} = 0.8{\bar{3}}$ throughout.
For boundary cases without polarity reversion see suppl.~fig.~\ref{fig:suppl:ODE_NoTurn}.
%%% Omit an extra panel for a heatmap $\tmo$ as function of $\delta$ and$\epsilon_n$, as it is very similar to that for the IPS
\textbf{D.} Time of minimal order $\tmoMF$ shown as contour lines (isotemporales) for the mean-field model (dashed, eq.~\eqref{eq:ODE_expressed_in_rR}) and the IPS (solid, same data as in fig.~\ref{fig:Results_IPS_size100}C,D), for levels 0.04...0.08, 0.10, 0.12, 0.15, 0.20. 
$\beta=1$.
\textbf{E.} For $\epsilon_n > 4.5$ and $\delta \searrow \delta_{crit} \left( \epsilon_n \right) > 0$, $\tmoMF$ grows faster than exponentially indicating a phase transition to non-turning dynamics (grey shaded region in bottom right of D). 
}
\label{fig:results_Mean-field}
\end{figure}

%0 general plan for the section
To unravel the quantitative dependence between the accelerating impact of $\delta$, the decelerating effect of $\epsilon_n$ and the role of the de-novo polarisation rate $\beta$ for the time of minimal order, we analyse the mean-field approximation.
We numerically solve the mean-field model~\eqref{eq:ODE_expressed_in_rR} in Morpheus \cite{Back2012,StarrussDeBack2014_Morpheus} using the Runge{--}Kutta discretisation scheme with time step~$10^{-6}$ and the initial configuration specified by eq.~\eqref{eq:initial_distribution}. 
The temporal evolution of the mean-field fractions~$\MFA{a}_i \left( t \right)$ of cells in state~$\mathbf{e}_i$ is sketched exemplarily in fig.~\ref{fig:results_Mean-field}A, and shown together with the observable~$\MFA{ \mathbf{p} }$ derived from suppl.~eq.~\eqref{suppl:representations_of_p_MF} in panels B and C, respectively.
%1 Always turning, except for $\epsilon_n =5$, $\delta=0$
For all parameter combinations studied in the IPS model, the polarity pattern described by the mean-field model, eq.~\eqref{eq:ODE_expressed_in_rR_a_0still_free}, reorganise within the simulated time window $t \in \left[0, t_{max} \right] = \left[0, 1\right]$ as well, reversing the main polarisation direction and adapting to the global signal, except for the border cases of $\delta = 0, \, \epsilon_n \in \left\{4.5, 5 \right\}$. 
The time courses of $ \left\| \MFA{ \mathbf{p} } \right\|$ and $\operatorname{ang} \left( \MFA{ \mathbf{p} } \right)$ for the turning cases exhibit several common characteristics independent of the specific parameter sets and with those for the original model, described in the supplement~\ref{SI:NumericalSolutionOfMean-fieldODE} in detail.
In particular, there is a distinct time of minimal order $\tmoMF$ which characterises the time requirement of polarity pattern reorientation in the mean-field model.
For the non-turning cases, the neighbour coupling strength $\epsilon_n$ is too high such that the mean-field approximation is no longer usable as discussed in sec.~\ref{subsubsec:mean-fieldAnalysis} and exemplarily shown in suppl.~fig.~\ref{fig:suppl:ODE_NoTurn}. 

%parts of 1, 2 and 3 moved to supplement

%4 parameter dependence for $t_mo$
\paragraph{Parameter dependence of time of minimal order $\tmoMF$ }
The measured time of minimal order $\tmoMF =  \tmoMF \left( \delta, \epsilon_n \right)$ from our numerical simulations of the non-linear mean-field model~\eqref{eq:ODE_expressed_in_rR} is reported in fig.~\ref{fig:results_Mean-field}D as isotemporales alongside with data for $\tmo$ from the IPS simulations. 
The two data sets coincide very well qualitatively and quantitatively. 
We conclude that the mean-field approximation is a suitable tool to study the polarity reorientation in the dynamically diluted alignment model throughout a wide parameter range, and that the qualitative interpretation of parameter dependencies extends from IPS (cf. section~\ref{subsec:Results:IPSdata}) to the mean-field approximation.
For border cases $\epsilon_n > 4.5$ and $\delta \searrow \delta_{crit}$ for critical value $ \delta_{crit} =  \delta_{crit} \left(\epsilon_n \right) > 0$, the time of minimal order $\tmoMF$ grows faster than exponentially as evidence of the transition to non-turning polarity patterns (fig.~\ref{fig:results_Mean-field}E). 

%5 analytical expression for epsilon_n = 0
For $\epsilon_n = 0$, cells evolve independently and the time of minimal order can be determined analytically from an ODE as 
\begin{align}   \label{eq:tmo_for_epsilon_n=0}
 \tmoMF \left( \epsilon_n =0 \right)   \quad = \quad   \frac{  \log 2  }{  \delta +   \sum_{k=1}^{8}  \exp \left( \epsilon_s \left\langle  \mathbf{s} , \mathbf{e}_k \right\rangle \right)   }  ,
\end{align}
see suppl.~eq.~\eqref{suppl:ODE_for_p_expressed_in_rR} in suppl.~\ref{SI:tmo_for_epsilon_n=0}.
Moreover, $\tmoMF \left( \epsilon_n =0 \right) $ coincides with time of minimal order for the averaged order parameter $\langle \mathbf{p} \rangle$, see suppl.~\ref{SI:tmo_for_epsilon_n=0} and suppl.~fig.~\ref{fig:suppl:IPSvsMF_epsilon_n=0}.

\subsection{Linearised mean-field ODE}  \label{subsec:LinearMean-fieldODE}
\begin{figure} 
{\includegraphics[width=1.0 \textwidth, height= \textheight, keepaspectratio]{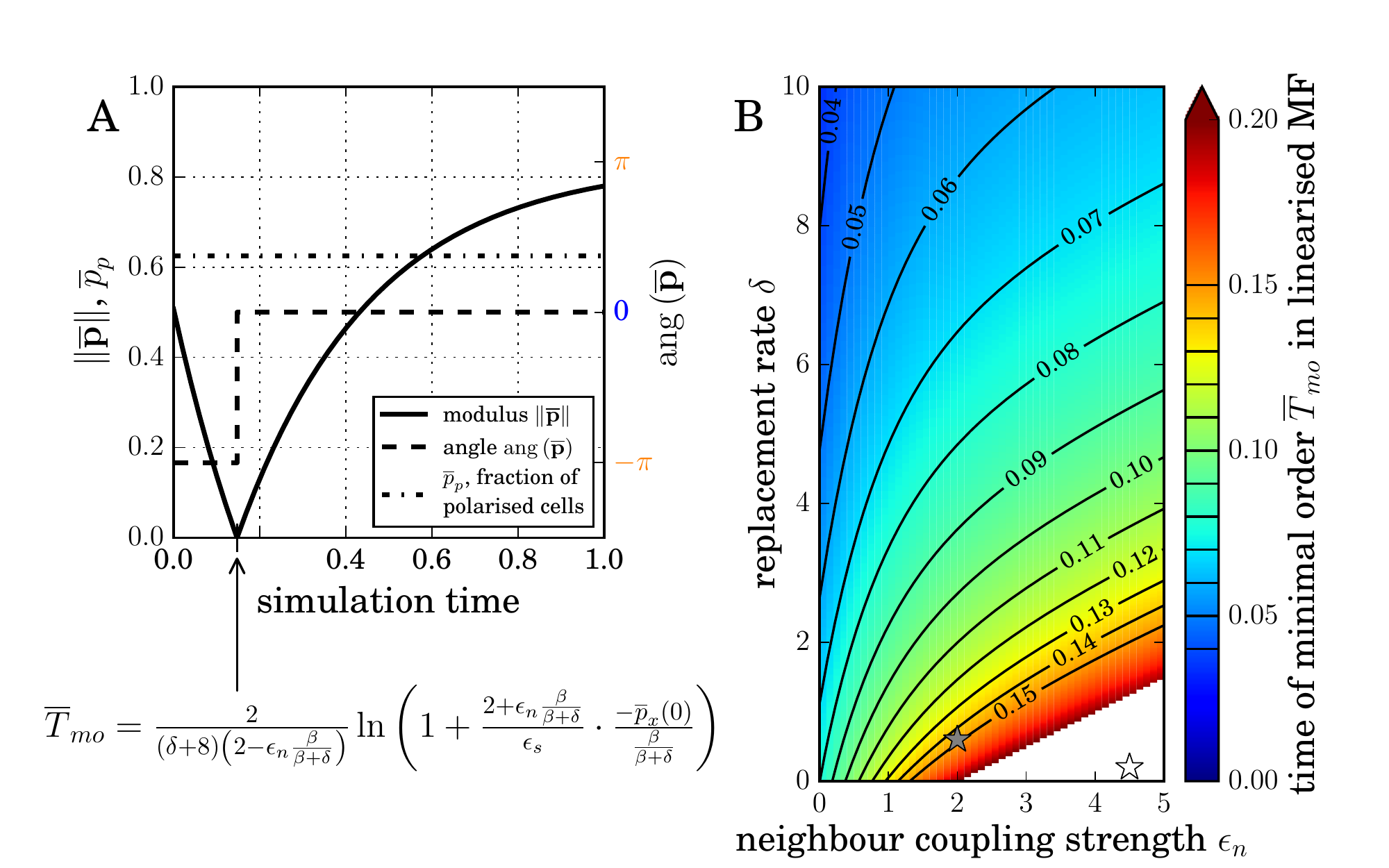}}
\caption{
Linearised mean-field model. 
\textbf{A.} Analytical solution of linear ODE for $\linMFA{p}_x \left(t\right)$ yields $\linMFA{ \mathbf{p} }$, cf.~suppl.~eqs. \eqref{suppl:linearised_ODE_for_p_x_p_y_decoupled_linMFa_0_steady}, \eqref{suppl:t_mo_linearised_MF_most_general}. 
Modulus $\left\| \linMFA{ \mathbf{p} } \right\|$ (solid, left axis) starts high, reaches a unique minimum with $\left\| \linMFA{ \mathbf{p} } \right\| = 0$ that defines $\tmolinMF$, and increases again to a high plateau.
The angle $\operatorname{ang} \left( \linMFA{ \mathbf{p} } \right)$ (dashed, right axis) switches from $-\pi$ to $0$ at $\tmolinMF$. 
The fraction of polarised cells (dash-dotted, left axis) remains constantly $\linMFA{p}_p = p_{eq} = 0.625$ throughout.
Parameters are $\epsilon_s=1$, $\beta = 1$, $\delta = 0.6$ and $\epsilon_n=2.0$ (denoted by grey asterisk in B and different from parameter values in figs.~\ref{fig:Results_IPS_size100}A,B and \ref{fig:results_Mean-field}A-C, white asterisk in B).
\textbf{B.} Time of minimal order $\tmolinMF$ from eq.~\eqref{eq:t_mo_linearised_MF_general} shown as heatmap with contourlines (isotemporales), $\beta=1$. Note divergence for $\epsilon_n \frac{\beta}{\beta + \delta} \nearrow 2$.
}
\label{fig:results_linearMean-field}
\end{figure}

%1 Argue for further simplification
Because of the good quantitative agreement in the time of minimal order between the dynamically diluted alignment model ($\tmo$) and its mean-field approximation ($\tmoMF$), we investigate the latter model further in the linearised mean-field form of eq.~\eqref{eq:linearisedODE}, an ODE system that allows analytical treatment. 
%2 Stability discussion
The stability behaviour of the solution for the order parameter ODE depends on $A \define  \left( \delta + 8 \right) \left(-1 +  \frac{\epsilon_n}{2} \frac{\beta}{\beta + \delta} \right)$, see suppl.~eqs.~\eqref{suppl:t_mo_linearised_MF_most_general} and~\eqref{suppl:linearisedODE}.
For $A < 0$ or equivalently $ \epsilon_n  \frac{\beta}{\beta + \delta} < 2 $, the solution converges to a unique stable equilibrium.
For $A > 0$ or equivalently $ \epsilon_n  \frac{\beta}{\beta + \delta} > 2 $, there is a unique but unstable equilibrium and the solution diverges.
The order parameter in steady state, whether stable or not,
\begin{align} \label{eq:p_eq_from_LinearisedODE}
{\linMFA{ \mathbf{p} }}^{\ast}  =  \left( - \frac{B_x}{A}, - \frac{B_y}{A} \right) = \frac{\beta}{\beta + \delta} \cdot  \frac{1}{2 - \epsilon_n  \frac{\beta}{\beta + \delta}}  \epsilon_s \mathbf{s}   
\end{align}
is aligned with the global signal~$\mathbf{s}$ in the stable, therefore convergent, case, but counter-directional in the unstable, therefore divergent, case.
%3 Rule out divergent case
This together with the fact that $\left\| {\linMFA{ \mathbf{p} }} \left( t \right) \right\|   \stackrel{t \to \infty}{ \longrightarrow}  \left\| {\linMFA{ \mathbf{p} }}^{\ast} \right\| > 1 $ for 
$ 2 - \epsilon_s \left\| \mathbf{s} \right\| \frac{\beta}{\beta + \delta}  <     \epsilon_n \frac{\beta}{\beta + \delta}   <    2 + \epsilon_s \left\| \mathbf{s} \right\| \frac{\beta}{\beta + \delta} $ 
in contrast to $\left\|  \mathbf{p} \left( t \right) \right\| \leq 1$ in the IPS
indicates that the linearised ODE system~\eqref{eq:linearisedODE} and derived equations (\eqref{eq:t_mo_linearised_MF_general}, suppl.~eqs.~\eqref{suppl:linearisedODE} and \eqref{suppl:linearised_ODE_for_p_x_p_y_decoupled_linMFa_0_steady}) are an appropriate approximation of the dynamically diluted alignment model only in the convergent case $ \epsilon_n  \frac{\beta}{\beta + \delta} < 2 $.
The divergent case is a spurious solution introduced by the errors of MFA and linearisation that both grow with $ \epsilon_n  \frac{\beta}{\beta + \delta}$.
%4 Results for convergent case suggest parameter rescaling
Fig.~\ref{fig:results_linearMean-field}B visualises the parameter dependence of $\tmolinMF $ for $\epsilon_s = 1$, $s_x = 1$ (cf. eq.~\eqref{eq:t_mo_linearised_MF_general}) and the initial condition given by eq.~\eqref{eq:initial_distribution}. 
The results coincide with $\tmo$ in the IPS description qualitatively and even quantitatively, except for divergence of $\tmolinMF$ for $ \epsilon_n  \frac{\beta}{\beta + \delta} \nearrow 2$ as evidence for the transition to the divergent case (compare fig.~\ref{fig:Results_IPS_size100}D to fig.~\ref{fig:results_linearMean-field}B).
Note that in eqs.~\eqref{eq:t_mo_linearised_MF_general}, \eqref{eq:p_eq_from_LinearisedODE} and suppl.~eq.~\eqref{suppl:t_mo_linearised_MF_most_general}, neighbour coupling strengths $\epsilon_n$ is rescaled with a factor $\frac{\beta}{\beta + \delta} $ that equals the fraction of polarised cells $p_{eq}$, cf.~eq.~\eqref{eq:frac_pol}, thereby modulating neighbour influences with cell turnover. 
This suggests to revisit the empirical results from the IPS simulations and non-linear MFA as a function of this parameter combination, which we call \emph{effective} neighbour coupling strength,
\begin{align} \label{eq:Def_eps_n_eff}
\epsilon_n^{eff}  \define  \frac{\beta}{\beta + \delta} \epsilon_n  .
\end{align}

\subsection{Analysis of simulation results through parameter rescaling}  \label{subsec:Results_rescaled}

\begin{figure}
\centering
\includegraphics[width=1.0 \textwidth, keepaspectratio]{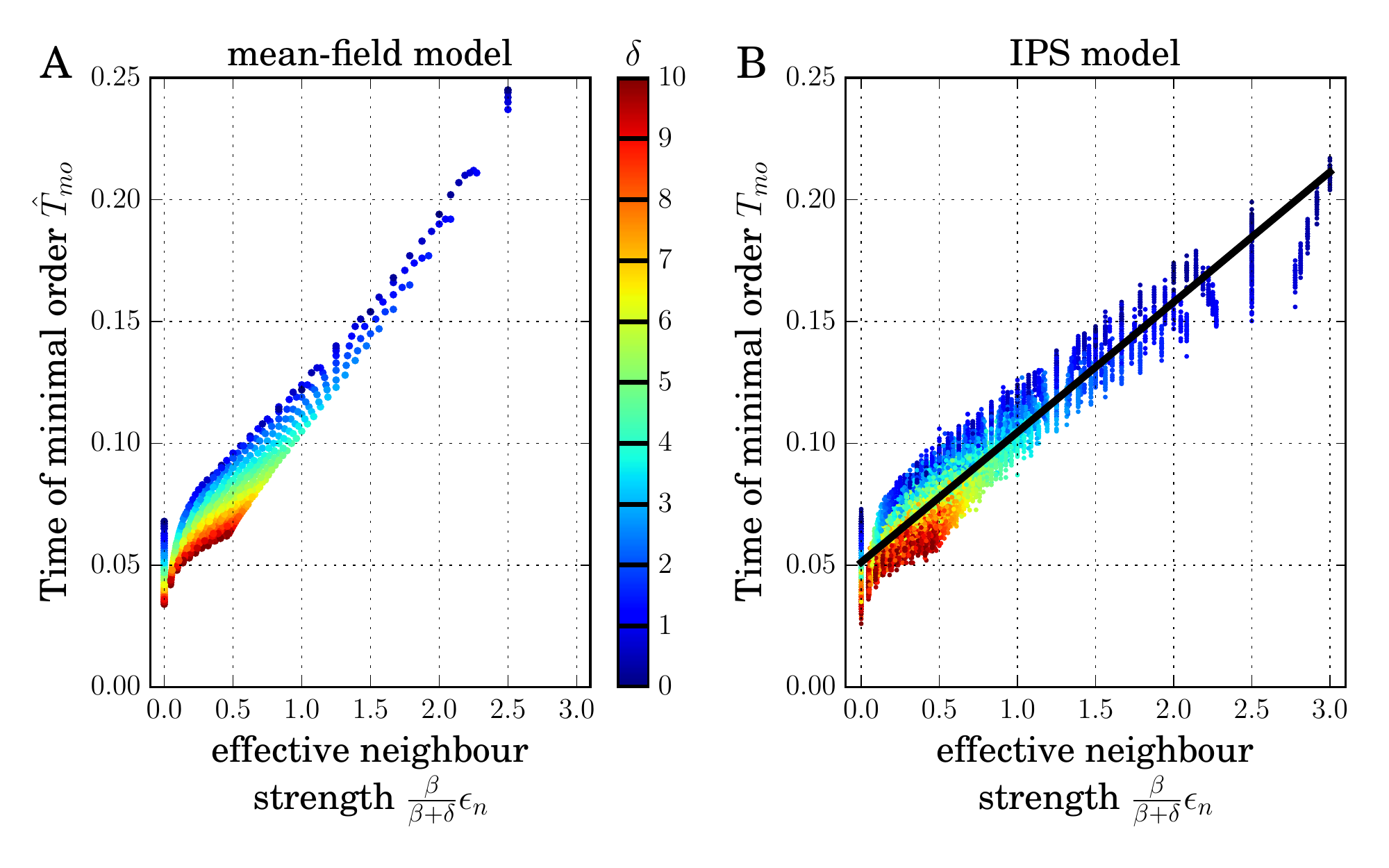}
\caption{
Measured collapse onto a linear dependence upon rescaling to \textit{effective} neighbour coupling strength. 
\textbf{A.} Mean-field model, same data as in fig.~\ref{fig:results_Mean-field}D, rescaled. Parameter $\delta$ is identically color coded in both panels. 
\textbf{B.} All data of IPS simulations (same data as in fig.~\ref{fig:Results_IPS_size100}C,D) collapse to eq.~\eqref{eq:tmo=functionOf_eps_n}, given by solid curve and grey belt. % of orthogonal distance regression % for 100x100 data
}
\label{fig:Result_TimeOfMinimalOrder_in_IPS_collapse}
\end{figure}

The identification of the effective neighbour coupling strength $\epsilon_n^{eff}$ in the analysis of the linearised mean-field model above suggests to rescale the simulation results according to this parameter combination.
Fig.~\ref{fig:Result_TimeOfMinimalOrder_in_IPS_collapse} shows the data from our stochastic IPS model (from fig.~\ref{fig:Results_IPS_size100}C,D) and from numerical solutions of the non-linear mean-field model (from fig.~\ref{fig:results_Mean-field}D) versus the rescaled parameter $\epsilon_n^{eff} = \frac{\beta}{\beta+\delta}\epsilon_n $. 
We find that all data of figs.~\ref{fig:Results_IPS_size100}C,D and \ref{fig:results_Mean-field}C collapse on to approximately linear relationships.
For the IPS model with $\epsilon_n = 1$, the linear relationship is 
\begin{align}   \label{eq:tmo=functionOf_eps_n}
\tmo = \tmo \left( \epsilon_n^{eff} \right) = (0.0533\pm0.0006) \underbrace{  \frac{\beta}{\beta+\delta}  }_{ =p_{eq} } \epsilon_n + (0.0512\pm0.0005) % for 100x100 data
\end{align}  % more  digits:   \tmo = \tmo \left( \epsilon_n^{eff} \right) = (0.05338643\pm0.00065415) \underbrace{  \frac{\beta}{\beta+\delta}  }_{ =p_{eq} } \epsilon_n + (0.05119605\pm0.0005141)
where values in brackets denote $(\textnormal{mean} \pm \textnormal{std} )$ of an orthogonal distance regression.
The inverse relationship reads 
\begin{align}   \label{eq:eps_n=functionOf_tmo}
\epsilon_n =  \bigl(  (18.7314\pm0.2295) \tmo  -   (0.9590\pm0.0200)  \bigr)  / \left( \frac{\beta}{ \beta + \delta } \right)
\end{align}   %more digits: \epsilon_n =  \bigl(  (18.731357\pm0.22951925 ) \tmo  +   ( -0.95897158\pm0.01978007)  \bigr)  / \left( \frac{\beta}{ \beta + \delta } \right)
with experimentally accessible variables $\tmo$ and $\frac{\beta}{ \beta + \delta } = p_{eq}$ on the right hand side. 

In the mean-field model, the critical value $\delta_{crit} \left( \epsilon_n \right)$ for the transition to non-turning dynamics is transformed into a critical effective neighbour coupling strength $\epsilon_{n, crit}^{eff} \approx 4.60$ valid throughout the considered parameter space. 
According to our simulation data and model analysis, the impact of cell replacement on tissue polarity reorganisation can be well described as weakening of the neighbours' influence on each single cell's polarisation by the fraction of unpolarised cells.
Altogether, neighbour coupling retards polarity pattern reorganisation, whereas cell turnover accelerates it. 
The time of minimal order and effective neighbour coupling strength are related by an approximately linear function.

\subsection{Determination of parameter values}
Our model analysis allows to estimate the actual parameter value of $\epsilon_n$ from experimental observations.
Eq.~\eqref{eq:eps_n=functionOf_tmo} only requires $p_{eq}$ and $\tmo$, that can be measured, 
and indirectly the time unit $1/\gamma$ which relates to the frequency of polarity changes, see eq.~\eqref{eq:Def_rate_pol_to_pol}.
To determine the latter, we see different options.
First, it might be obtained using time-lapse imaging of sub-cellular polarity markers.
Second, analysis of calibrated models of the molecular processes underlying tissue polarity may provide estimates for $\gamma$.
Third and most realistically, the dimensionless eq.~\eqref{eq:tmo=functionOf_eps_n} can be written as 
\begin{align}   \label{eq:tmo=functionOf_eps_n_in_timeUnits}
\tmo =  \tilde T_{mo}  \frac{1}{ \tilde \gamma } =  (0.0533\pm0.0006)  p_{eq}  \epsilon_n + (0.0512\pm0.0005) 
\end{align}  
where $\tilde T_{mo}$ denotes the value \emph{with} physical time unit.
Measuring $\tilde T_{mo}$ in experiments for constant neighbour coupling strength $\epsilon_n$ and at least two different values of $p_{eq} =  \frac{\beta}{\beta+\delta} $, the instances of eq.~\eqref{eq:tmo=functionOf_eps_n_in_timeUnits} form a linear equation system that can be solved for $ \left( \gamma , \epsilon_n \right)$.
Variations of cell death rate~$\delta$ and/or de-novo polarisation rate~$\beta$ are feasible using drugs, RNAi and other techniques that interfere with pathways of cell proliferation and apoptosis.
The absolute value of the neighbour coupling strength $\epsilon_n$, obtained through any of the above ways, can then be interpreted in relation to the coupling strength to the global cue, here chosen as $\epsilon_s = 1$. 

Beyond such interpretation of absolute values, the comparison of inferred values for  $\epsilon_n$ among multiple perturbed conditions within a screening approach allows to disentangle the mechanistic effects of perturbations on neighbour coupling versus cell turnover versus intracellular polarity dynamics. Our theoretical results have resolved how all three contributions jointly determine the time scale of polarisation reorientation. As our proposed protocol only requires to measure quantities that are accessible from still images, this theory alleviates the need for life imaging of the same specimen.

\section{Discussion}

% 1. recall question and results
% 1a. question
We have here posed the question how local and global instructing signals are integrated in a planar cell polarity system.
In particular, conflict resolution as observed in double-headed planaria is proposed as a valuable source of information about which signals dominate the temporal evolution that are not accessible from studying random initial conditions.
To study this problem theoretically, we propose a cell-based IPS model accounting for local cell-cell coupling of polarity, for sensitivity to global ligand gradients and for the impact of cell turnover as present in planaria, termed dynamically diluted alignment model. 

% 1b results IPS
Analysing the model numerically and analytically, we find that the global signal dictates the final tissue polarisation orientation independent of the strength of neighbour coupling between cells and the amount of cell turnover.
The temporal evolution from an initially polarised tissue conflicting with the global signal to the final polarised state in accordance with the global signal occurs via a disordered state in which each polarisation direction is approximately equally abundant.
We introduce this time of minimal order~$\tmo$ as an observable to measure the time scale of conflict resolution and study its dependency on cell turnover rate and neighbour coupling strength.
It turns out that neighbour coupling retards polarity pattern reorganisation whereas cell turnover accelerates it, and that dependencies are gradual without abrupt transitions.
We employ mean-field analysis of the IPS model to derive an ODE for the temporal evolution of the average polarisation and an equation for the time of minimal order which well approximate the IPS data. 
As a result, we identify an effective neighbour coupling strength which integrates the parameters of cell turnover and neighbour coupling, %cell turnover: \delta AND \beta
and demonstrate that the time of minimal order in the dynamically diluted alignment model depends linearly on the effective neighbour coupling strength.

% 2 Interpretation
% 2a Model
% 2a 0 dynamically diluted alignment model captures cell turnover
The dynamically diluted alignment model developed here extends former approaches where the effects of neighbour coupling on PCP polarity establishment were studied by means of a cell-grained tissue polarity model \cite{BurakShraiman2009}.
Our model accounts for cell turnover which is present in most biological tissues.
From theoretical point of view, dynamical site dilution is a generalisation of static site dilution as studied in statistical physics in the context of ferromagnetism.
Annealed site dilution of Potts models has been formulated previously, but studies so far focussed on steady state properties rather than temporal dynamics.
Moreover, planar cell polarity is a widespread phenomenon in live matter and provides an experimental realisation of an annealed site-dilution Potts model.

% 2a 1 . biological implications
% reduce signals to $s$
In the model, we neglect the molecular details of specific pathways underlying cell polarity reorganisation, because we are interested in polarity conflict resolution between cellular and tissue scales and aim for a model which is still analytically tractable. 
In planaria, the molecular details of the PCP pathway and its upstream global signals are getting unraveled but their role for planar tissue polarity needs to be studied further \cite{AdellCebriaSalo2010,Almuedo2011,Brooks2016,Azimzadeh2016,Stueckemann2017}.
%In other species, including \textit{Xenopus} and mice, direct links have been established between the core PCP pathway and cilia orientation, beating or gliding direction \cite{Wallingford2010,Vladar2012}.  
For our model, all contributing tissue scale signals were subsumed into the abstract global director~$\mathbf{s}$.
In the same spirit, the orienting cues of the local cell-cell coupling are summarised as average neighbour polarisation vector.
Independent of the specific molecular pathways, cell-cell bridging complexes do not move across three-cell junctions but are degraded at one cell interface and assembled anew at another interface.
Therefore, we deliberately do not consider gradual changes in polarisation direction, but model the change of cell polarity direction as independent of the current polarisation direction of the considered cell.
The resulting model is equally well applicable to study Frizzled/Flamingo- or Fat/Dachsous-based patterning or other mechanisms of tissue polarity at cellular resolution in biological tissues with and without cell turnover.

% 2a 1b. other mechanisms that could reorient PCP incl. tissue shear (moved from Introduction)
In addition to the molecular cues, tissue polarity patterns can respond to mechanical shear of the cell packing, stemming from external forces or oriented cell divisions \cite{AigouyEtAlEaton2010,Heisenberg2013}.
In the case of a pre-existing planar tissue polarity pattern and mechanical tissue rearrangements due to oriented cell divisions in the direction of planar tissue polarity and inheritance of the mother cell's PCP pattern by the daughter cells, the tissue elongates but the planar tissue polarity pattern is maintained \cite{RodrigoAlbors2015}. 
On the other hand, if oriented cell divisions occur at an oblique angle to the direction of planar tissue polarity, then the orchestrated turning of pre-existing planar tissue polarity is observed as the result of induced tissue shear, both experimentally in the developing fly wing and in numerical model simulations \cites{AigouyEtAlEaton2010}{VichasZallen2011}. 
% Other modes of cell turnover, related to tissue shear
However, oriented cell divisions as observed to reorganise PCP in the fly wing play no role for planaria and tissue shear is negligible \cite{ReddienAlvarado2004}.  \\ 
% 2b. technical choices and limitations of our model and analysis
The IPS model follows the inherent discretisation of tissue into cells and describes the dynamics in continuous time as a stochastic process, which reflects noise and randomness in molecular interactions underlying the polarity patterns.
It was described here for a square lattice with von-Neumann neighbourhood but the model definition is valid as well for other lattices, like hexagonal or even image-derived lattices, and for more general neighbourhood templates. 
We expect similar results for other lattice geometries since the mean-field approximation, which is independent of the actual spatial arrangement, closely agrees with the original model.
A model extension which includes cell migration and cell division is straightforward \cite[e.g.][]{Voss-Boehme2010,Talkenberger2017underReview,Lee1995,Simpson2007},
but their effects on tissue polarity patterns have not been in the focus of our study.
The number of polarised states is an implicit model parameter. 

Our choice of eight polarisation vectors is the smallest number that conforms with the four-fold rotational symmetry of the lattice and allows, besides perpendicular and opposing polarity directions, also partial alignment with a directing polarity signal.
Also, it is known that the planar $q$-Potts model without external field exhibits a first order phase transition only for $q>4$, whereas the case $q=4$ can be reduced to $q=2$, i.e. the Ising model \cite{Wu1982}.
Note that by considering an ordered initial condition and a conflicting external field in this work, the emergence of spontaneous order and corresponding phase transitions cannot be studied.
From the analytical point of view, the choice of more than four cell polarisation states also contributes to the good agreement between IPS data and mean-field approximation.
In the limit $q\to\infty$, the $q$-Potts model yields the XY-model with continuous angle space and the Beresinkii{--}Kosterlitz{--}Thouless transition \cite{Kosterlitz1973}. 
However, we don't consider this limit an appropriate model for polarity patterns in epithelia since there are a number of three-cell junctions around each cell where transmembrane protein complexes cannot form. 
Hence a continuous polarity angle is only possible within finite angle ranges that are separated by narrow excluded angle ranges. 
The transient dynamics of such a piecewise continuous model may on short time scales resemble that of the XY-model but on long time scales that of the Potts model.
% 2c global signal dominates
Concerning the results of our analysis on the asymptotic steady state, we remark that the observed universal dominance of the global signal on the long-term tissue polarisation direction is in agreement with the known behaviour of Potts models \cite{Wu1982}.
However, the detailed temporal dynamics and the question of which signal dominates in a conflicting situation in the context of planar tissue polarity have not been studied before.
This indicates planar tissue polarity as a versatile experimental framework that represents Potts models with dynamic site-dilution.

% 2d \tmo is function of effective neighbour coupling strength
Given the knowledge that the global signal determines the long-term behaviour of planar tissue polarity, it is plausible that 
neighbour coupling retards polarity pattern reorganisation since it enforces the maintenance of the initial, locally coherent but globally conflicting direction.
For the same reason cell replacement has an accelerating effect since it facilitates resolution of contradictory signals by reducing the number of polarised neighbours.
By quantifying that $\tmo$ depends linearly on the effective neighbour coupling strength $\epsilon_n^{eff} = \frac{\beta}{\beta + \delta} \epsilon_n $ (fig.~\ref{fig:Result_TimeOfMinimalOrder_in_IPS_collapse}), our theory enables the estimation of the neighbour coupling strength from eq.~\eqref{eq:eps_n=functionOf_tmo} or \eqref{eq:tmo=functionOf_eps_n_in_timeUnits} by measuring~$\tmo$ and the fraction of polarised cells~$p_{eq}=\frac{\beta}{\beta + \delta}$.
This allows to infer the relative importance of global versus local directing signals, and to predict the effects of altering cell turnover on the time scale of tissue polarity reorganisation.
In particular in the finite time frame of an experiment, it is possible that modified cell turnover prolongs the time required for reorganisation beyond the observation window.

% 2e MF method
% 2e1. results MF
Mean-field approximation of each cell's local director field by the average of all fields allows further analytical results, notably the derivation of $\epsilon_n^{eff}$, but neglects spatial correlations that become important for high neighbour coupling strength~$\epsilon_n$.
This discrepancy induces a phase transition in the mean-field results that is not observed in our dynamically diluted alignment model itself. 
However, the artificially introduced phase transition occurs not until $\epsilon_n^{eff} \approx 4.6$, making mean-field approximation a valuable tool that provides qualitative and quantitative match for a wide range of parameters.
% 2e2. results linearised MF
The linearised mean-field model as a further simplification (eq.~\eqref{eq:linearisedODE}) provides an analytical expression for the polarity state and the time of minimal order (eq.~\eqref{eq:t_mo_linearised_MF_general}) but shifts the spurious phase transition down to $\epsilon_n^{eff} = 2$.
% 2e3 extensions
The mean-field analysis can be improved when the independence assumption is replaced by a kind of pair approximation \cite{Lucas2012}. 
This can reduce approximation errors and extend the range of applicability towards higher neighbour coupling strength $\epsilon_n$, for which the mean-field results under the independence assumption so far deviate from the IPS simulation results.

% 5. suggestions for extensions, outlook
The proposed model and its analysis performed here are ready to be applied to quantitative data. Direct or indirect measurements of tissue polarity with cellular resolution in \textit{S.~mediterranea} would allow to determine the model parameters, since quantified local alignment of polarity directions as a time and space dependent order parameter or decay of correlations directly link experimental data to observables of the model. The determined parameter set then implies further model predictions that could be tested experimentally.
The model is also applicable to discriminate between several hypothetical ligands that might provide the global signal. Provided they have different spatio-temporal concentration profiles, these can be tested \textit{in silico} to reproduce the observed tissue polarity reorientation patterns.

\section{Acknowledgements, Authors' Contributions, Funding and Competing Interests}
{\small{
% 1 Acknowledgement
The authors are grateful to Hanh Thi-Kim Vu for help with imaging planaria and acknowledge fruitful discussions with Walter de Back, Michael K{\"u}cken, J{\"o}rn Starru{\ss} and Carsten Timm.
% 2 funding
This work was supported by the German Federal Ministry of Education and Research (BMBF) under funding codes 0316169 and 031L0033.
AVB acknowledges support by S{\"a}chsisches Staatsministerium f{\"u}r Wissenschaft und Kunst (SMWK) in the framework of INTERDIS-2.
The simulations were performed on HPC resources granted by the ZIH at TU Dresden.
% 3 Author contributions
The study was designed by KBH, AVB and LB. 
JCR designed the experiments.
KBH performed model simulations and analysis, and generated figures.
KBH, AVB, JCR and LB interpreted the data, wrote and revised the manuscript.
% 4 conflicts of interest
We have no competing interests.
}}

\printbibliography

%====================================================================================================================
% SUPPLEMENT

% change names of the float environments
\addto\captionsenglish{\renewcommand{\contentsname}{Contents of Supplement}}
\addto\captionsenglish{\renewcommand{\figurename}{Supplementary Figure}}
\addto\captionsenglish{\renewcommand{\tablename}{Supplementary Table}}
\addto\captionsenglish{\renewcommand{\listfigurename}{List of Supplementary Figures}}
\addto\captionsenglish{\renewcommand{\listtablename}{List of Supplementary Tables}}
% number supplemental material with preceding 'S'
\renewcommand{\thesection}{S\arabic{section}}
\renewcommand{\thefigure}{S\arabic{figure}}
\renewcommand{\thetable}{S\arabic{table}}
\renewcommand{\theequation}{S\arabic{equation}}
% reset numbering of section, equations etc.
\setcounter{section}{0}
\setcounter{figure}{0}
\setcounter{table}{0}
\setcounter{equation}{0}

\newpage

\hspace*{5cm}

\begin{center}
\begin{LARGE}
Supplement to: \\

A Dynamically Diluted Alignment Model Reveals the Impact of Cell Turnover on the Plasticity of Tissue Polarity Patterns
\end{LARGE}
\end{center}

\newpage
%\tableofcontents
%\listoftables
%\listoffigures

\begin{table*}[h]
\begin{tabular}{llc}
property & symbol(s) \& remarks & \\  
\hline 
$\mathbf{e}_0 = \left(0,0\right)$   &  polarisation state, unpolarised &   \\ 
$\mathbf{e}_i  = \left(\cos\left(i\pi/4\right),\sin\left(i\pi/4\right)\right)$  &   polarisation state, polarised, see main text eq.~\eqref{eq:Def_e_i}, $i = 1,2,\mathellipsis,8$ &   \\ 
$\pmb{\nu}_z = \pmb{\nu}_z \left(\pmb{\eta} \right) \define \frac{ \sum_{x \in N_z} \pmb{\eta}_x  }{ \# N_z }  $     &  average neighbour direction in node $z \in S$  & \\   
$N_z$  & neighbour nodes of $z \in S$ in von Neumann neighbourhood &   \\
$\pmb{\nu} = \pmb{\nu} \left( \pmb{\eta} \right) = \left( \pmb{\nu}_z \left(\pmb{\eta} \right) \right)_{z \in S}$   &   field of average neighbour directions &       \\ 
$\epsilon_n$  &  neighbour coupling strength (sensitivity to neighbour polarisations)   &    \\  
$\mathbf{s}$  &   vector of global signal&     \\   
$\epsilon_s$  &  coupling strength to the global signal (sensitivity to global signal $\mathbf{s}$)   &       \\  
$\mathbf{w}_z \define \epsilon_n \pmb{\nu}_z \left( \pmb{\eta} \right) + \epsilon_s \mathbf{s}$  &  reference orientation, see  main text eq.~\eqref{eq:Def_w} and main text fig.~\ref{fig:ModelDef}  &    \\
$\gamma$ & overall rate of reorientation; unit model time is set to $1 / \gamma$  \\    
$\delta$   &  loss-of-polarisation rate (cell death rate)    &      \\
 $\beta$ &  de novo polarisation rate    &     \\ 
$\mathbf{p} = \mathbf{p} \left( \pmb{\eta} \right)  \define  \frac{ \sum_{z \in S} \pmb{\eta}_z }{  \# S }$  &  (globally) average polarisation in IPS, see main text eq.~\eqref{eq:Def_orderParam}  &      \\
$ \tmo $  &  time of minimal order in IPS  &    \\
$ a_i \left( \pmb{\eta} \right)  =  \frac{ \# \left\{ z \in S, \, \pmb{\eta}_z = \mathbf{e}_i \right\}  }{ \# S  }   $  &  fraction of nodes in polarisation state $i$ in IPS, see main text eq.~\eqref{eq:Def_a_i}, \\
 & \qquad  $i=0, \mathellipsis , 8$ &   \\
$p_p = 1 -  \frac{\# \left\{ z \in S; \, \eta_z = \mathbf{e}_0 \right\} }{\# S} $  &  fraction of polarised cells in IPS, see main text eq.~\eqref{eq:Def_p_p}  &    \\
$ p_{eq} = \frac{ \beta }{ \beta + \delta }$  &  equilibrium fraction of polarised cells, see eq.~\eqref{eq:frac_pol}   &     \\
$ \epsilon_n^{eff} = \frac{ \beta }{ \beta + \delta } \epsilon_n $  &  effective neighbour coupling strength, see main text eq.~\eqref{eq:Def_eps_n_eff}   &       \\
$  \tmoMF,  \MFA{ \mathbf{p} }, \MFA{ \mathbf{a} }, \MFA{a}_i, \MFA{p}_p, \mathellipsis $  &  properties under mean-field approximation  &   \\
$  \tmolinMF, \linMFA{ \mathbf{p} }, \linMFA{ \mathbf{a} }, \linMFA{a}_i, \linMFA{p}_p, \mathellipsis $  &  properties under mean-field approximation and linearisation  &   \\
\hline
\end{tabular}
\caption[List of symbols.]{List of symbols. The number of elements of a set $A$ is denoted $\# A $.}
\label{tab:ListOfSymbols}
\end{table*}

\section{Description of Supplementary Movie}  \label{suppl:movie_beta=1}

The movie \textrm{'DynamicallyDilutedAlignmentModel\_SupplementaryMovie.mpeg'},
contained in the electronic supplementary material, 
shows the trajectory of a typical simulation of the IPS model on a $100 \times 100$~lattice with periodic boundaries, $\epsilon_n=4.5$, $\delta=0.2$, $\beta=1$, $\epsilon_s=1$, $\mathbf{s}=\left(1,0\right)$. 
The movie covers 1 time unit in the dedimensionalised model time; time resolution of visualisation is $0.001$.
See main text fig.~\ref{fig:ModelDef}C for colour code.
The bottom left detail of the same simulation is shown in main text fig.~\ref{fig:Results_IPS_size100}A for times~0.05,~0.2,~0.5,~0.8 (\textbf{a-d}),
and its analysis is shown in main text fig.~\ref{fig:Results_IPS_size100}B.

\newpage

\section{Details of the mean-field analysis}  \label{suppl:mean-fieldAnalysis}

Using mean-field approximation, we derive an ODE which approximates the temporal evolution of $\mathbf{p} \left( t \right) = \mathbf{p} \left( \pmb{\eta} \left( t \right) \right)$.
Denote the fraction of nodes in states $\mathbf{e}_0, \mathellipsis , \mathbf{e}_8$ at time $t$ in the IPS model by 
\begin{align}  \label{suppl:Def_a_i}
a_i \left( t \right)  \define  a_i \left( \pmb{\eta} \left( t \right) \right)  \define \frac{ \# \left\{ z \in S, \, \pmb{\eta}_z \left( t \right) = \mathbf{e}_i \right\}  }{\# S }  , \qquad i=0, \mathellipsis , 8 
\end{align}
as in main text eq.~\eqref{eq:Def_a_i}.
Writing shorthand $\mathbf{a} \left( t \right) \define \left(a_1 \left( t \right), \mathellipsis , a_8 \left( t \right) \right)^{\transpose}$ where $\left( { } \cdot { } \right)^{\transpose}$ denotes matrix transposition, the mean polarisation vector is related to vector $\mathbf{a}$ via main text eq.~\eqref{eq:representations_of_p}
\begin{align}  \label{suppl:representations_of_p}
\mathbf{p} \left( t \right) = \left( p_x \left( t \right), \, p_y \left( t \right) \right)^{\transpose} =  \sum_{i=1}^{8} a_i \left( t \right) \cdot \mathbf{e}_i = M \mathbf{a} \left( t \right)
\end{align}
where
\begin{align}  \label{suppl:Def_MatrixM} 
M \define  \begin{pmatrix}  \frac{1}{2} \sqrt{2} & 0 & -\frac{1}{2} \sqrt{2} & -1 & -\frac{1}{2} \sqrt{2} & 0 & \frac{1}{2} \sqrt{2} & 1   \\   \frac{1}{2} \sqrt{2} & 1 & \frac{1}{2} \sqrt{2} & 0 & -\frac{1}{2} \sqrt{2} & -1 & -\frac{1}{2} \sqrt{2} & 0   \end{pmatrix}  .
\end{align}
Note that $ a_0 \left( t \right) = 1 - \sum_{i=1}^{8} a_i \left( t \right)$ is determined by $\mathbf{a} \left( t \right)$ and that the fraction of polarised cells is $p_p \left( t \right) = 1- a_0 \left( t \right)$. 

The mean-field assumption (MFA) presumes that the local director field acting at a single node can be approximated by the average field of all nodes. 
Hence the local director  $\pmb{\nu}_z \left( \pmb{\eta} \right)$, see main text eq.~\eqref{eq:Def_nu}, is approximated by
\begin{align}   \label{suppl:MFA_nu}
\pmb{\nu}_z \left( \pmb{\eta} \left( t \right) \right)   =  \frac{1}{ \# N_z }  \sum_{x \in N_z}  \pmb{\eta}_x    \left( t \right)   \stackrel{MFA}{\approx}  \frac{1}{ \#S }  \sum_{x \in S} \pmb{\eta}_x   \left( t \right) = \mathbf{p}  \left( t \right)    .
\end{align}
Then main text eq.~\eqref{eq:Def_rate_pol_to_pol} becomes
\begin{align}  \label{suppl:MFA_rate_pol_to_pol}    
\operatorname{c}_z  \left( \pmb{\eta} , \mathbf{e}_i \right)  & \stackrel{\hphantom{MFA}}{=}  \exp \left(  \left\langle \mathbf{e}_i, \, \epsilon_n \cdot \pmb{\nu}_z \left(\pmb{\eta} \right) + \epsilon_s \cdot \mathbf{s} \right\rangle \right)    \nonumber  \\
 { } & \stackrel{MFA}{\approx}    \exp \left(  \left\langle \mathbf{e}_i, \,  \epsilon_n \cdot \mathbf{p} + \epsilon_s \cdot \mathbf{s}, \right\rangle \right)   \nonumber   \\
 { } & \stackrel{\hphantom{MFA}}{=}    \exp \left( \left\langle \mathbf{e}_i, \,  \epsilon_n \cdot M \mathbf{a} + \epsilon_s \cdot \mathbf{s} \right\rangle \right) \enifed  r_i \left( \mathbf{a} \right) ,     \qquad \textnormal{if } \pmb{\eta}_z \neq \mathbf{e}_0, \, i=1,\mathellipsis,8      ,
\end{align}
therewith introducing the substitutes $r_i \left( \mathbf{a} \right)$.
Analogously, main text eq.~\eqref{eq:Def_rate_unpol_to_pol} is approximated as
\begin{align} \label{suppl:MFA_rate_unpol_to_pol}    
\operatorname{c}_z  \left( \pmb{\eta} ,  \mathbf{e}_i \right)  & \stackrel{\hphantom{MFA}}{=}  \beta \cdot \frac{\exp \left(  \left\langle \mathbf{e}_i, \, \epsilon_n \cdot \pmb{\nu}_z \left(\pmb{\eta} \right) + \epsilon_s \cdot \mathbf{s} \right\rangle \right) }{ \sum_{k=1}^{8} \exp \left(  \left\langle \mathbf{e}_k, \, \epsilon_n \cdot \pmb{\nu}_z \left(\pmb{\eta} \right) + \epsilon_s \cdot \mathbf{s} \right\rangle \right) }     \nonumber    \\
 { } &  \stackrel{MFA}{\approx}     \beta \cdot \frac{\exp \left(  \left\langle \mathbf{e}_i, \, \epsilon_n \cdot \mathbf{p} + \epsilon_s \cdot \mathbf{s} \right\rangle \right) }{ \sum_{k=1}^{8} \exp \left(  \left\langle \mathbf{e}_k, \, \epsilon_n \cdot \mathbf{p} + \epsilon_s \cdot \mathbf{s} \right\rangle \right) }   \nonumber    \\
 { } &  \stackrel{\hphantom{MFA}}{=}  \beta \frac{r_i \left( \mathbf{a} \right)} {\sum_{k = 1}^{8}  r_k \left( \mathbf{a} \right) }  ,   \qquad \textnormal{if } \pmb{\eta}_z  = \mathbf{e}_0 , \, i =1,\mathellipsis, 8      .
\end{align}
For convenience, we abbreviate 
\begin{align} \label{suppl:Def_R}
R \left( \mathbf{a} \right) \define \sum_{ k =1 }^{8} r_{k} \left( \mathbf{a} \right)   ,
\end{align}
such that
\begin{align}  \label{suppl:MFA_rate_unpol_to_pol_shorter}
\operatorname{c}_z  \left( \pmb{\eta} ,  \mathbf{e}_i \right)  \stackrel{MFA}{\approx}  \beta \, \frac{r_i \left( \mathbf{a} \right)} {  R \left( \mathbf{a} \right)  }    ,   \qquad \textnormal{if } \pmb{\eta}_z  = \mathbf{e}_0 , \, i =1,\mathellipsis, 8      
\end{align}
as stated in main text eq.~\eqref{eq:MFA_rate_pol_to_pol_AND_unpol_to_pol}.
Note that the approximations in eqs.~\eqref{suppl:MFA_rate_pol_to_pol}, \eqref{suppl:MFA_rate_unpol_to_pol} and \eqref{suppl:MFA_rate_unpol_to_pol_shorter} are exact for $\epsilon_n = 0$ and that the approximation error increases with $\epsilon_n$.

In the limit for increasing lattice size the $a_i$'s become continuous quantities and their dynamic behaviour can be described by an ODE system [references 48--50 in the main text].
%\cite{vanKampen1997,Boettger2015,Hohmann2013}.
For convenience, main text eq.~\eqref{eq:ODE_expressed_in_rR_a_0still_free} is reproduced here:
\begin{align} \label{suppl:ODE_expressed_in_rR_a_0still_free} 
\left.    \begin{aligned}  
    \frac{\mathrm{d} \MFA{a}_i }{\mathrm{d} t }   &  =   - \MFA{a}_i  \cdot   \left( \delta + \sum_{ k =1 }^{8} r_k \left( \MFA{ \mathbf{a} } \right)  \right)    +    \MFA{a}_0 \cdot \beta \frac{r_i \left(  \MFA{ \mathbf{a} } \right)} {\sum_{k = 1}^{8}  r_k \left(  \MFA{ \mathbf{a} } \right)}  +   \sum_{ k =1}^{8}  \MFA{a}_k \cdot r_i \left(  \MFA{ \mathbf{a} } \right)      \\   
    { }   &  =   - \MFA{a}_i  \cdot   \left( \delta + R \left(  \MFA{ \mathbf{a} } \right)  \right) +  \left[ \MFA{a}_0 \cdot \frac{\beta}{R \left(  \MFA{ \mathbf{a} } \right)} + \left(1-\MFA{a}_0\right)   \right] r_i \left(  \MFA{ \mathbf{a} } \right)   , \qquad i=1,\mathellipsis, 8      \\
    \frac{\mathrm{d} \MFA{a}_0 }{\mathrm{d} t }  &   = - \MFA{a}_0 \cdot \beta + \delta \cdot \left(1 - \MFA{a}_0 \right)  =   \delta  - \left(\beta + \delta  \right) \MFA{a}_0      .     
\end{aligned}     \right\}    
\end{align}
Here $\MFA{ \mathbf{a} }$ and $\MFA{a}_i$ denote the counterparts of $ \mathbf{a}$ and $a_i$ under mean-field approximation (MFA).
We call eq.~\eqref{suppl:ODE_expressed_in_rR_a_0still_free} the \emph{mean-field model} and note that the overall error of approximation introduced in eqs.~\eqref{suppl:MFA_rate_pol_to_pol} and \eqref{suppl:MFA_rate_unpol_to_pol} increases with $\epsilon_n$, $\epsilon_s$, and the fraction of polarised cells~$p_p$.
Note that other, even irregular, lattice geometries, different neighbourhood templates and spatially asymmetrically weighted neighbour polarity information yield the same MFA \cref{suppl:ODE_expressed_in_rR_a_0still_free} as long as neighbour polarity information is weighted independently of the considered cell's polarity state.

The fraction of unpolarised cells $\MFA{a}_0 \left( t \right)$ decouples with unique solution 
\begin{align}    \label{suppl:ODE_dead_fraction}
 \MFA{a}_0 (t)  =  \frac{\delta }{ \beta + \delta } + \left(\MFA{a}_0 (0) - \frac{\delta }{ \beta + \delta } \right) \cdot \exp \bigl( - \left(\beta + \delta  \right) t \bigr)   , 
\end{align}
that tends to the unique, globally attracting equilibrium $\MFA{a}_0^{\ast} = \frac{\delta}{\beta + \delta} $. 
Hence the fraction of polarised cells converges, $\MFA{p}_p = 1 - \MFA{a}_0 \stackrel{t \to \infty}{\longrightarrow} \frac{\beta}{\beta + \delta}$, in perfect agreement with the dynamic equilibrium $p_{eq} = \frac{\beta}{\beta + \delta} $ of death and de novo polarisation in the original IPS (eq.~\eqref{eq:frac_pol}). 
We will exploit this steady state expressionfor $a_0$ and $\MFA{a}_0$ in various places. 
Inserting $\MFA{a}_0 = \MFA{a}_0^{\ast} = \frac{\delta}{\beta + \delta} $, the ODE system~\eqref{suppl:ODE_expressed_in_rR_a_0still_free} simplifies to main text eq.~\eqref{eq:ODE_expressed_in_rR}
\begin{align}  \label{suppl:ODE_expressed_in_rR}
\frac{\mathrm{d} \MFA{a}_i }{\mathrm{d} t }  &  =  - \MFA{a}_i   \cdot   \left( \delta + R \left( \MFA{ \mathbf{a} } \right) \right)  +     \frac{\beta}{\beta + \delta}  \cdot  r_i \left( \MFA{ \mathbf{a} } \right)  \cdot  \left( 1+  \frac{\delta}{ R \left( \MFA{ \mathbf{a} } \right) }    \right)    \nonumber  \\
 { } & =   \left( \delta + R \left( \MFA{ \mathbf{a} } \right) \right) \cdot \left(  -\MFA{a}_i  +   \frac{\beta}{\beta + \delta}  \frac{ r_i \left( \MFA{ \mathbf{a} } \right) }{ R \left( \MFA{ \mathbf{a} } \right) }   \right)  , \qquad i=1,\mathellipsis, 8    .  
\end{align}
An approximate solution for $\mathbf{p}  \left( t \right) = \mathbf{p} \left( \pmb{\eta} \left( t \right) \right)$, denoted $\MFA{ \mathbf{p} } \left( t \right)$ in the following,
can be obtained by solving the mean-field model ODE system~\eqref{suppl:ODE_expressed_in_rR} and using the mean-field analogue of eq.~\eqref{suppl:representations_of_p} (main text eq.~\eqref{eq:representations_of_p})
\begin{align}   \label{suppl:representations_of_p_MF}
\MFA{ \mathbf{p} }  \left( t \right) = \sum_{k=1}^{8}  \MFA{a}_k  \left( t \right) \cdot \mathbf{e}_k .
\end{align}
Alternatively, eqs.~\eqref{suppl:ODE_dead_fraction} and \eqref{suppl:ODE_for_p_expressed_in_rR} describe an ODE system. Equation 
\begin{align} \label{suppl:ODE_for_p_expressed_in_rR}
\frac{\mathrm{d} \MFA{ \mathbf{p} } }{\mathrm{d} t } \left( t \right)    & =   - \left( \delta + R \left( \MFA{ \mathbf{p} } \right) \right) \cdot \MFA{\mathbf{p}}    +     \left[ \MFA{a}_0 \cdot \frac{\beta}{R \left(  \MFA{ \mathbf{p} } \right)} + \left(1-\MFA{a}_0\right)   \right] \sum_{i=1}^8  r_i \left(  \MFA{ \mathbf{p} } \right) \cdot \mathbf{e}_i 
\end{align}
follows from summing \eqref{suppl:ODE_expressed_in_rR} for $i=1, \mathellipsis, 8$ and replacing $M \MFA{ \mathbf{a} }$ by $ \MFA{ \mathbf{p} }$ in \eqref{suppl:MFA_rate_pol_to_pol} to define $ r_i \left(  \MFA{ \mathbf{p} } \right)$ and $R \left( \MFA{ \mathbf{p} } \right)$.

We observe that solutions of \eqref{suppl:ODE_expressed_in_rR} preserve the symmetry of the initial condition with respect to the \mbox{$x$-axis}, that was imposed by setting $\mathbf{s}=\left(1,0\right)$, $a_1 \left(0\right) = a_7 \left(0\right)$, $a_2 \left(0\right) = a_6 \left(0\right)$ and $a_3 \left(0\right) = a_5 \left(0\right)$ (cf.~main text eq.~\eqref{eq:initial_distribution}). 
For such a symmetric initial condition, it holds that $\MFA{ p }_y \left( t \right) = 0$ for all $t \geq 0$ such that  $\MFA{ \mathbf{p} }  \left( t \right) = \left( \MFA{ p }_x \left( t \right), 0 \right) $.
We solve eq.~\eqref{suppl:ODE_expressed_in_rR} numerically in main text section~\ref{subsec:NumericalSolutionOfMean-fieldODE} and calculate $\left\| \MFA{ \mathbf{p} } \right\| = \left| \MFA{ p }_x \right|$ and $\operatorname{ang} \left( \MFA{ \mathbf{p} } \right) = \pi \cdot \operatorname{sgn} \left( \MFA{ p }_x \right)$.

The precise initial condition $\MFA{ \mathbf{a} }  \left( 0 \right)$ used for numerical solving of eq.~\eqref{suppl:ODE_expressed_in_rR} (main text eq.~\eqref{eq:ODE_expressed_in_rR}) follows from main text eq.~\eqref{eq:initial_distribution} as 
{\scriptsize{
\begin{align*}
\MFA{ a }_k \left( 0 \right) = \frac{\beta}{\beta + \delta}  \cdot  \frac{        \exp \left\{  \left( \epsilon_n+\epsilon_s \right)  \left\langle  \mathbf{e}_k ,  \,  \mathbf{e}_4  \right\rangle  \right\}   }{        \exp \left( \epsilon_n+\epsilon_s \right) +2 \exp \left( \sqrt{0.5} \left(\epsilon_n+\epsilon_s \right) \right)  + 2 +  2 \exp \left( - \sqrt{0.5} \left(\epsilon_n+\epsilon_s \right) \right)  +  \exp \left( - \left( \epsilon_n+\epsilon_s \right) \right)      }  ,   \\
  \qquad  k=1, \mathellipsis , 8    .
\end{align*}    
}}

\section{Linearisation of mean-field model}  \label{suppl:ODELinearisation}  
We can approximate the non-linear mean-field model (main text eq.~\eqref{eq:ODE_expressed_in_rR}, or equivalently eq.~\eqref{suppl:ODE_expressed_in_rR}) further to obtain an analytically tractable ODE.
Linearisation of $\exp \left( { } \cdot { } \right) $ in eq.~\eqref{suppl:MFA_rate_pol_to_pol} by Taylor expansion around~$0$ yields
\begin{align*}  
r_i \left( \mathbf{a} \right) & =  1 +  \epsilon_n  \sum_{k=1}^{8}   a_k \left\langle \mathbf{e}_k, \, \mathbf{e}_i \right\rangle + \epsilon_s \left\langle \mathbf{s}, \, \mathbf{e}_i \right\rangle + \mathcal{O}\left( \left( \epsilon_n + \epsilon_s \right)^2 \right)     , \qquad i=1,\mathellipsis, 8     ,
\end{align*} 
where the remainder term has been estimated using the bounds $\left| \left\langle \mathbf{e}_k, \, \mathbf{e}_i \right\rangle  \right| \leq 1 $,  $\left| a_k \right| \leq 1$ for $i,k =1, \mathellipsis, 8$. 
Inserting this result into eq.~\eqref{suppl:Def_R} one obtains
\begin{align*}  
R \left( \mathbf{a} \right) & =  \sum_{i=1}^{8} \left(   1 +  \epsilon_n  \sum_{k=1}^{8}   a_k \left\langle \mathbf{e}_k, \, \mathbf{e}_i \right\rangle + \epsilon_s \left\langle \mathbf{s}, \, \mathbf{e}_i \right\rangle + \mathcal{O}\left( \left( \epsilon_n + \epsilon_s \right)^2 \right)   \right)   \\
 { } & = 8 +  \epsilon_n  \sum_{k=1}^{8}   a_k \left\langle \mathbf{e}_k, \, \underbrace{ \sum_{i=1}^{8} \mathbf{e}_i }_{= \left(0,0\right)} \right\rangle  +  \epsilon_s \left\langle \mathbf{s}, \,  \underbrace{\sum_{i=1}^{8}  \mathbf{e}_i }_{= \left(0,0\right)} \right\rangle  + \mathcal{O}\left( \left( \epsilon_n + \epsilon_s \right)^2 \right)    \\
 { } & = 8 +  \mathcal{O}\left( \left( \epsilon_n + \epsilon_s \right)^2 \right)             .
\end{align*} 
To indicate this second approximation by linearisation, id est dropping $\mathcal{O}\left( \left( \epsilon_n + \epsilon_s \right)^2 \right)$ terms for $\left( \epsilon_n + \epsilon_s \right)^2 \ll 1$, we add an overline to the approximated quantities.
The ODE system~\eqref{suppl:ODE_expressed_in_rR} (or main text eq.~\eqref{eq:ODE_expressed_in_rR}) with linearised $r_i \left( { } \cdot { } \right)$'s reads then
\begin{align}   \label{suppl:linearisedODE}
\frac{\mathrm{d} \linMFA{a}_i }{\mathrm{d} t }   &  =   \left( \delta + 8 \right)  \left(    - \linMFA{a}_i  +  \frac{1}{8} \frac{\beta}{\beta + \delta}  \left[  1 +  \epsilon_n  \sum_{k=1}^{8}   \linMFA{a}_k \left\langle \mathbf{e}_k, \, \mathbf{e}_i \right\rangle + \epsilon_s \left\langle \mathbf{s}, \, \mathbf{e}_i \right\rangle   \right]  \right) , \qquad i=1,\mathellipsis, 8     ,
\end{align} 
which shows main text eq.~\eqref{eq:linearisedODE}. 

Using $  \linMFA{p}_x =   \left\langle \linMFA{ \mathbf{p} } , \mathbf{e}_8 \right\rangle$  with $  \linMFA{ \mathbf{p} } =  \sum_{k=1}^{8} \linMFA{a}_i  \cdot \mathbf{e}_i  $, 
we obtain (see extra suppl.~\ref{suppl:subsec:derive_maintext_equation} for detailed calculations)
\begin{align} \label{suppl:linearised_ODE_for_p_x_p_y_decoupled_linMFa_0_steady} 
 \frac{\mathrm{d} \linMFA{ p }_x }{\mathrm{d} t }  &  = \left( \delta + 8 \right) \left[ \left(-1 +  \frac{\epsilon_n}{2} \frac{\beta}{\beta + \delta} \right)  \linMFA{ p }_x +  \frac{\beta}{\beta + \delta}  \frac{\epsilon_s}{2} s_x \right] \, .
\end{align}
The linear ODE~\eqref{suppl:linearised_ODE_for_p_x_p_y_decoupled_linMFa_0_steady} has the form $ \frac{\mathrm{d} \linMFA{ p }_x \left( t \right) }{\mathrm{d} t }  =  A \linMFA{ p }_x + B_x$ 
with $A = \left( \delta + 8 \right) \left(-1 +  \frac{\epsilon_n}{2} \frac{\beta}{\beta + \delta} \right) $ and $B_x = \left( \delta + 8 \right) \frac{\beta}{\beta + \delta}  \frac{\epsilon_s}{2} s_x $ and hence has the solution 
\begin{align}  \label{suppl:t_mo_linearised_MF_most_general}
\linMFA{ p }_{x} \left( t \right) =  \left[  \linMFA{ p }_{x} \left( 0 \right)  + \frac{B_x}{A} \right] \exp \left( A t \right) - \frac{B_x}{A} \, .
\end{align}
Since the initial condition and parameter $\mathbf{s}$ are chosen symmetric with respect to the \mbox{$x$-axis} such that $ \linMFA{ p }_{y} \left( 0 \right) = 0$ and $s_y = 0$, it also holds that $\linMFA{ p }_y \left( t \right) = 0, \, t \geq t_0$ and  $\linMFA{ \mathbf{p} }  \left( t \right) = \left( \linMFA{ p }_x \left( t \right), 0 \right) $ and the symmetry is preserved over time.
Then equation~\eqref{suppl:linearised_ODE_for_p_x_p_y_decoupled_linMFa_0_steady} provides a simple ODE for $\linMFA{ \mathbf{p} }  \left( t \right)$ compared to deriving $\MFA{ \mathbf{p} }   \left( t \right) = \sum_{k=1}^{8} \MFA{a}_k \left( t \right) \mathbf{e}_k$ according to \eqref{suppl:representations_of_p} from a solution of~\eqref{suppl:ODE_expressed_in_rR} (main text eq.~\eqref{eq:ODE_expressed_in_rR}).
For non-symmetric initial conditions and more general $\mathbf{s}$, $\linMFA{ p }_y \left( t \right)$ will differ from $0$ and be governed by an ODE analogous to~\eqref{suppl:linearised_ODE_for_p_x_p_y_decoupled_linMFa_0_steady} with $x$ being replaced by $y$. 

When $p_x \left( 0 \right) $ opposes $s_x$, then $p_x \left( \cdot \right)$ must change sign for polarity reorientation. 
This occurs if and only if $A<0$ or equivalently $\epsilon_n \frac{\beta}{\beta + \delta} < 2$.
Then the time of minimal order in the linear MF model $\tmolinMF$ is given as the unique root of $p_x \left( \cdot \right) $ as
\begin{align}  \label{suppl:t_mo_linearised_MF_general}
\tmolinMF & =  \frac{1}{A} \ln \left(  \frac{B_x}{ A \cdot \linMFA{ p }_x  \left( 0 \right)+ B_x} \right)   \notag  \\    
   { } & =  \frac{2}{ \left( \delta + 8 \right) \left( 2 - \epsilon_n \frac{\beta}{\beta + \delta} \right)}   \ln \left( 1 +   \frac{2 + \epsilon_n \frac{\beta}{\beta + \delta}}{\epsilon_s}
          \cdot   \frac{ - \linMFA{ p }_x  \left( 0 \right)}{ s_x \frac{\beta}{\beta + \delta}}  \right)     ,
\end{align} 
which shows main text eq.~\eqref{eq:t_mo_linearised_MF_general}.
The last term $\frac{ - \linMFA{ p }_x  \left( 0 \right)}{ s_x \frac{\beta}{\beta + \delta}}$ is always positive due to our initial condition where $ \linMFA{ p }_x  \left( 0 \right)$ and $s_x$ have opposite signs.
Additionally, the term equals 1 in case of perfect alignment among polarised cells in the initial configuration.
In the case $\epsilon_n \frac{\beta}{\beta + \delta} > 2$ and hence $A>0$, $\linMFA{ p }_{x}$ does not change sign.
To summarise, polarity reverses in the linearised mean-field model only for $\epsilon_n \frac{\beta}{\beta + \delta} < 2$.

The precise initial condition $\linMFA{ p }_x  \left( 0 \right)$ in eq.~\eqref{suppl:t_mo_linearised_MF_general} (main text eq.~\eqref{eq:t_mo_linearised_MF_general}) follows from main text eq.~\eqref{eq:initial_distribution} as 
{\scriptsize{
\begin{align*}
\linMFA{ p }_x \left( 0 \right) = \frac{\beta}{\beta + \delta}  \cdot  \frac{             - \exp \left( \epsilon_n+\epsilon_s \right) - \sqrt{2} \exp \left( \sqrt{0.5} \left(\epsilon_n+\epsilon_s \right) \right)  +  \sqrt{2} \exp \left( - \sqrt{0.5} \left(\epsilon_n+\epsilon_s \right) \right)  +  \exp \left( - \left( \epsilon_n+\epsilon_s \right) \right)        }{        \exp \left( \epsilon_n+\epsilon_s \right) +2 \exp \left( \sqrt{0.5} \left(\epsilon_n+\epsilon_s \right) \right)  + 2 +  2 \exp \left( - \sqrt{0.5} \left(\epsilon_n+\epsilon_s \right) \right)  +  \exp \left( - \left( \epsilon_n+\epsilon_s \right) \right)      }  .     
\end{align*}    
}}

\section{Detailed derivation of supplement equation \eqref{suppl:linearised_ODE_for_p_x_p_y_decoupled_linMFa_0_steady}  } \label{suppl:subsec:derive_maintext_equation}

Recall the preceding eq.~\eqref{suppl:linearisedODE}.
By an analog of eq.~\eqref{suppl:representations_of_p} (or main text eq.~\eqref{eq:representations_of_p}) holds $  \linMFA{ \mathbf{p} } \left( t \right) = \left( \linMFA{p}_x \left( t \right), \, \linMFA{p}_y \left( t \right) \right)^{\transpose} = \sum_{i=1}^{8} \linMFA{a}_i  \left( t \right)  \cdot \mathbf{e}_k$.
Hence with $\mathbf{e}_8 = \left(1, \, 0 \right)$
\begin{align*}
 \linMFA{p}_x \left( t \right)  & = \left\langle \linMFA{ \mathbf{p} } \left( t \right) , \mathbf{e}_8 \right\rangle  \\
 { } & = \sum_{i=1}^{8} \linMFA{a}_i  \left( t \right)  \cdot \left\langle \mathbf{e}_i , \mathbf{e}_8 \right\rangle     . 
\end{align*}
Plugging in eq.~\eqref{suppl:linearisedODE} yields
\begin{align}   \label{suppl:eq:linearisedODE_plugged_in}
 \frac{\mathrm{d} \linMFA{ p }_x }{\mathrm{d} t }  \left( t \right) & =   \sum_{i=1}^{8}  \frac{\mathrm{d}  \linMFA{a}_i }{\mathrm{d} t }  \left( t \right)  \cdot \left\langle \mathbf{e}_i  , \mathbf{e}_8 \right\rangle    \notag \\
 { } & =  \left( \delta + 8 \right)  \sum_{i=1}^{8} \left[ \left(    - \linMFA{a}_i  \left( t \right)  +  \frac{1}{8} \frac{\beta}{\beta + \delta}  \left[  1 +  \epsilon_n  \sum_{k=1}^{8}   \linMFA{a}_k \left( t \right) \left\langle \mathbf{e}_k, \, \mathbf{e}_i \right\rangle + \epsilon_s \left\langle \mathbf{s}, \, \mathbf{e}_i \right\rangle   \right]  \right) \cdot \left\langle \mathbf{e}_i , \mathbf{e}_8 \right\rangle  \right]    \notag  \\
 { } &  =    - \left( \delta + 8 \right)  \left\langle \sum_{i=1}^{8}  \linMFA{a}_i  \left( t \right) \cdot \mathbf{e}_i , \mathbf{e}_8 \right\rangle     +    \left( \delta + 8 \right) \frac{1}{8} \frac{\beta}{\beta + \delta}   \sum_{i=1}^{8} \left\langle \mathbf{e}_i , \mathbf{e}_8 \right\rangle    \notag  \\    
 { } & \qquad  +     \left( \delta + 8 \right) \frac{1}{8} \frac{\beta}{\beta + \delta}  \epsilon_n   \sum_{i=1}^{8} \sum_{k=1}^{8} \linMFA{a}_k \left( t \right) \left\langle \mathbf{e}_k, \, \mathbf{e}_i \right\rangle  \left\langle \mathbf{e}_i , \mathbf{e}_8 \right\rangle   \notag   \\   
 { } &  \qquad +    \left( \delta + 8 \right) \frac{1}{8} \frac{\beta}{\beta + \delta}  \epsilon_s   \sum_{i=1}^{8}  \left\langle \mathbf{s}, \, \mathbf{e}_i \right\rangle  \left\langle \mathbf{e}_i , \mathbf{e}_8 \right\rangle  
\end{align}
Note that $\sum_{i=1}^{8} \left\langle \mathbf{e}_i , \mathbf{e}_8 \right\rangle  =  \left\langle  \sum_{i=1}^{8} \mathbf{e}_i , \mathbf{e}_8 \right\rangle = 0$.
Further, 
\begin{align*}
\sum_{i=1}^{8} \sum_{k=1}^{8} \linMFA{a}_k \left( t \right) \left\langle \mathbf{e}_k, \, \mathbf{e}_i \right\rangle  \left\langle \mathbf{e}_i , \mathbf{e}_8 \right\rangle   &   = \sum_{i=1}^{8} \left\langle \sum_{k=1}^{8} \linMFA{a}_k \left( t \right) \mathbf{e}_k, \, \mathbf{e}_i \right\rangle  \left\langle \mathbf{e}_i , \mathbf{e}_8 \right\rangle   \\
 { } & =  \sum_{i=1}^{8} \left\langle \linMFA{ \mathbf{p} } \left(t \right), \, \mathbf{e}_i \right\rangle  \left\langle \mathbf{e}_i , \mathbf{e}_8 \right\rangle   .
\end{align*}
Because of the choice of unit vectors $\mathbf{e}_i  = \left(\cos\left(i\pi/4\right),\sin\left(i\pi/4\right)\right)$, $i = 1,2,\mathellipsis,8$, cf.~main text eq.~\eqref{eq:Def_e_i}, holds for arbitrary vector $\mathbf{v} = \left( v_x, v_y \right) \in \mathbb{R}^2 $
\begin{align*}
\sum_{i=1}^8  \left\langle \mathbf{v}, \mathbf{e}_i \right\rangle  \left\langle \mathbf{e}_i, \mathbf{e}_8 \right\rangle   & =    \sum_{i=1}^8  \left\langle \mathbf{e}_i,  v_x \mathbf{e}_8 + v_y \mathbf{e}_2  \right\rangle  \left\langle \mathbf{e}_i, \mathbf{e}_8 \right\rangle   \\
 { } & =    \sum_{i=1}^8  \left\langle \mathbf{e}_i, \mathbf{e}_8 \right\rangle ^2  v_x  +    \sum_{i=1}^8  \left\langle \mathbf{e}_i, \mathbf{e}_8 \right\rangle   \left\langle \mathbf{e}_i,   \mathbf{e}_2  \right\rangle  v_y   \quad = \quad  4 v_x      .
\end{align*}
We employ this identity specifically for $\mathbf{v} = \left( v_x, v_y \right) \in \left\{ \mathbf{s} ,  \linMFA{ \mathbf{p} } \left(t \right)  \mid t \in \left[ 0, \infty \right) \right\}$. Together with the aforementioned relations this simplifies eq.~\eqref{suppl:eq:linearisedODE_plugged_in} to 
\begin{align*}
 \frac{\mathrm{d} \linMFA{ p }_x }{\mathrm{d} t }  \left( t \right) & = - \left( \delta + 8 \right)  \left\langle \linMFA{ \mathbf{p} } \left( t \right) , \mathbf{e}_8 \right\rangle    +     \left( \delta + 8 \right) \frac{1}{8} \frac{\beta}{\beta + \delta}  \epsilon_n \cdot 4 \linMFA{ p }_x \left(t \right)    +    \left( \delta + 8 \right) \frac{1}{8} \frac{\beta}{\beta + \delta}  \epsilon_s  \cdot  4 s_x   \\ 
 { } & =  \left( \delta + 8 \right)  \left[ - \linMFA{ p }_x \left(t \right)  +  \frac{4}{8} \frac{\beta}{\beta + \delta}  \epsilon_n \linMFA{ p }_x \left(t \right)     +   \frac{4}{8} \frac{\beta}{\beta + \delta}  \epsilon_s   s_x  \right]
\end{align*}
and yields the desired equation~\eqref{suppl:linearised_ODE_for_p_x_p_y_decoupled_linMFa_0_steady}
\begin{align*}
\frac{\mathrm{d} \linMFA{ p }_x }{\mathrm{d} t }  &  = \left( \delta + 8 \right) \left[ \left(-1 +  \frac{\epsilon_n}{2} \frac{\beta}{\beta + \delta} \right)  \linMFA{ p }_x +  \frac{\beta}{\beta + \delta}  \frac{\epsilon_s}{2} s_x \right] 
\end{align*}
as claimed.   \qed

\section{Derivation of main text equation \eqref{eq:tmo_for_epsilon_n=0} for vanishing neighbour coupling strength $\epsilon_n = 0$}  \label{SI:tmo_for_epsilon_n=0}

For vanishing neighbour coupling strength $\epsilon_n = 0$, cells evolve independently as is evident from the definition of the IPS rates in eqs.~\eqref{eq:Def_rate_pol_to_pol}, \eqref{eq:Def_rate_pol_to_unpol} and \eqref{eq:Def_rate_unpol_to_pol}.
In particular, rates do only depend on the global signal~$\mathbf{s}$ and the state of the cell itself (polarised or not), but not on neighbour cells.
Hence, mean-field approximation in eqs.~\eqref{suppl:MFA_rate_pol_to_pol}, \eqref{suppl:MFA_rate_unpol_to_pol} and \eqref{suppl:MFA_rate_unpol_to_pol_shorter} (or main text eq.~\eqref{eq:MFA_rate_pol_to_pol_AND_unpol_to_pol}) is exact.
Moreover, the auxiliary rates $r_i \left( \mathbf{a} \right)$, $R  \left( \mathbf{a} \right)$ defined in eqs.~\eqref{suppl:MFA_rate_pol_to_pol} and \eqref{suppl:Def_R}, respectively, become independent of the actual fraction of cells in each of the nine states as
\begin{align}  \label{suppl:eq:Rr_for_epsilon_n=0}
r_i  &  = \exp \left(  \epsilon_s \left\langle \mathbf{e}_i, \mathbf{s} \right\rangle \right)    , \quad   i =1,\mathellipsis,8,  \\
R & =  \sum_{i=1}^8 \exp \left(  \epsilon_s \left\langle \mathbf{e}_i, \mathbf{s} \right\rangle \right)  .
\end{align}

Hence eq.~\eqref{suppl:ODE_for_p_expressed_in_rR} simplifies to
\begin{align} \label{suppl:eq:ODE_for_p_expressed_in_rR_for_epsilon_n=0}
\frac{\mathrm{d} \MFA{ \mathbf{p} } }{\mathrm{d} t } \left( t \right)    & =   - \left( \delta + R \right) \cdot \MFA{\mathbf{p}}    +     \left[ \MFA{a}_0 \cdot \frac{\beta}{R } + \left(1-\MFA{a}_0\right)   \right] \sum_{i=1}^8  r_i  \cdot \mathbf{e}_i 
\end{align}
and with steady state of cell death and de novo polarisation ($\MFA{a}_0 = \MFA{a}_0^{\ast} = \frac{\delta}{\beta + \delta} $, confer eq.~\eqref{suppl:ODE_dead_fraction}) further to
\begin{align} \label{suppl:eq:ODE_for_p_expressed_in_rR_for_epsilon_n=0_a_0=steady}
\frac{\mathrm{d} \MFA{ \mathbf{p} } }{\mathrm{d} t } \left( t \right)    & =   - \left( \delta + R \right) \cdot \MFA{\mathbf{p}}    +     \frac{\beta}{\beta + \delta} \frac{\delta + R}{R} \sum_{i=1}^8  r_i  \cdot \mathbf{e}_i 
\end{align}
which is a linear ODE for $\MFA{\mathbf{p}} $.
Note that $ \left( \delta + R \right)$ and $  \frac{\beta}{\beta + \delta} \frac{\delta + R}{R}$ are scalar factors in this 2-dimensional ODE.
For simpler notation write the right=most term as $\left( \delta + R \right) \cdot \mathbf{q} $ where $ \mathbf{q} \define  \frac{\beta}{\beta + \delta} \frac{1}{R} \sum_{i=1}^8  r_i  \cdot \mathbf{e}_i $ .
Then the general solution of \eqref{suppl:eq:ODE_for_p_expressed_in_rR_for_epsilon_n=0_a_0=steady} reads
\begin{align} \label{suppl:eq:p_solved_for_epsilon_n=0_a_0=steady}
 \MFA{ \mathbf{p} }  \left( t \right)    & =   \exp \left\{ - \left( \delta + R \right) \left( t - t_0 \right) \right\} \left[   \MFA{ \mathbf{p} }  \left( t_0 \right)    -   \mathbf{q}  \right]   +   \mathbf{q}   .
\end{align}
Let $t_0 = 0$. To find the minima of the order parameter $\left\| \MFA{ \mathbf{p} }  \left( t \right)  \right\| $ one can equally consider its square, that can be expressed in terms of scalar products as
\begin{align*} 
 \left\| \MFA{ \mathbf{p} }  \left( t \right)  \right\| ^2   & =  \left\langle  \MFA{ \mathbf{p} } \left( t \right) , \, \MFA{ \mathbf{p} } \left( t \right) \right\rangle \\
  { }  & = \exp \left\{ - 2 \cdot \left( \delta + R \right) \cdot t  \right\}   \left\|  \MFA{ \mathbf{p} }  \left( 0 \right)  - \mathbf{q} \right\|^2  +    2  \exp \left\{ - \left( \delta + R \right) \cdot t  \right\}  \left\langle  \MFA{ \mathbf{p} }  \left( 0 \right)  - \mathbf{q} , \,  \mathbf{q}  \right\rangle   +   \left\|   \mathbf{q} \right\|^2          .
\end{align*}
The first and last summands are non-negative.
Hence, if $\left\langle  \MFA{ \mathbf{p} }  \left( 0 \right)  - \mathbf{q} , \,  \mathbf{q}  \right\rangle   \geq  0$ the modulus $ \left\| \MFA{ \mathbf{p} }  \left( t \right)  \right\| $ decreases monotonically with time $t$ towards the limit value $\left\|   \mathbf{q} \right\|$. In this case no distinct time of minimal order exists.
In contrast, if $\left\langle  \MFA{ \mathbf{p} }  \left( 0 \right)  - \mathbf{q} , \,  \mathbf{q}  \right\rangle   < 0$ then there is a unique time of minimal order $\tmoMF \left(\epsilon_n =0 \right)$ that we find from 
\begin{align*}
 0  & =  \frac{\mathrm{d} \left(  \left\| \MFA{ \mathbf{p} }  \left( t \right)  \right\| ^2   \right)}{ \mathrm{d} t}  \left( \tmoMF \right)  \\
 { } & = - 2 \cdot \left( \delta + R \right)  \exp \left\{ - 2 \cdot \left( \delta + R \right) \cdot  \tmoMF \right\}   \left\|  \MFA{ \mathbf{p} }  \left( 0 \right)  - \mathbf{q} \right\|^2    \\
 { } & \qquad \qquad \qquad    - 2  \cdot \left( \delta + R \right)   \exp \left\{ - \left( \delta + R \right) \cdot  \tmoMF  \right\}  \left\langle  \MFA{ \mathbf{p} }  \left( 0 \right)  - \mathbf{q} , \,  \mathbf{q}  \right\rangle 
\end{align*} 
as
\begin{align} \label{suppl:eq:tmo_for_epsilon_n=0_general}
\tmoMF \left( \epsilon_n =0 \right)  & = \frac{1}{ \delta + R }  \cdot \log \left( - \frac{\left\|  \MFA{ \mathbf{p} }  \left( 0 \right)  - \mathbf{q} \right\|^2}{\left\langle  \MFA{ \mathbf{p} }  \left( 0 \right)  - \mathbf{q} , \,  \mathbf{q}  \right\rangle}  \right)   .
\end{align} 

We now determine $\MFA{ \mathbf{p} }  \left( 0 \right)$ for the initial condition described by main text eq.~\eqref{eq:initial_distribution}.
Note that $\mathbf{e}_k = - \mathbf{e}_{\left(k+4\right) \mod 8}$ from the defining eq.~\eqref{eq:Def_e_i} and rewrite the initial condition (main text eq.~\eqref{eq:initial_distribution}) as 
\begin{align*} 
 \MFA{ \mathbf{p} }  \left( 0 \right)    & =   \frac{ \beta }{ \left( \beta + \delta \right) } \frac{ 1 }{ Z } \sum_{k=1}^8 \exp \left(  \epsilon_s  \left\langle \mathbf{e}_k, \,  \mathbf{e}_4  \right\rangle  \right) \mathbf{e}_k     \\
  { } & =  \frac{ \beta }{ \left( \beta + \delta \right) } \frac{ 1 }{ Z } \sum_{k=1}^8 \exp \left(  \epsilon_s  \left\langle \mathbf{e}_{\left(k+4\right) \mod 8}, \,  \mathbf{e}_8  \right\rangle  \right) \mathbf{e}_{\left(k+4\right) \mod 8}     \\ 
  { } & =  \frac{ \beta }{ \left( \beta + \delta \right) } \frac{ 1 }{ Z } \sum_{i=1}^8 \exp \left(  \epsilon_s  \left\langle \mathbf{e}_i , \,  \mathbf{e}_8  \right\rangle  \right) \cdot \left( - \mathbf{e}_i \right)   \\ 
  { } & =  - \frac{ \beta }{ \left( \beta + \delta \right) } \frac{ 1 }{ Z } \sum_{i=1}^8 r_i \mathbf{e}_i   
\end{align*}
The normalisation denominator is $Z = \sum_{i=1}^8 r_i = R$, such that 
\begin{align} \label{suppl:eq:initial_distribution_for_epsilon_n=0}
 \MFA{ \mathbf{p} }  \left( 0 \right)    & =  - \frac{ \beta }{ \left( \beta + \delta \right) } \frac{  \sum_{i=1}^8 r_i \mathbf{e}_i }{R}  & = - \mathbf{q}  .
\end{align}
Hence $\left\langle  \MFA{ \mathbf{p} }  \left( 0 \right)  - \mathbf{q} , \,  \mathbf{q}  \right\rangle  =  \left\langle  - \mathbf{q}  - \mathbf{q} , \,  \mathbf{q}  \right\rangle  = - 2 \left\| \mathbf{q}  \right\|^2  < 0$, so there is a uniquely determined time of minimal order\footnote{
We neglect the case $ \mathbf{q} = \mathbf{0}$, which is only possible for $\epsilon_s = 0$, in addition to the assumption of $\epsilon_n = 0$.
 }
$\tmoMF$. Plugging \cref{suppl:eq:initial_distribution_for_epsilon_n=0} into \cref{suppl:eq:tmo_for_epsilon_n=0_general} yields
\begin{align*}
\tmoMF \left( \epsilon_n \right) = \frac{1}{\delta +  R } \cdot \log \left( - \frac{\left\|  - \mathbf{q}   - \mathbf{q} \right\|^2}{\left\langle - \mathbf{q}  - \mathbf{q} , \,  \mathbf{q}  \right\rangle}  \right)  =   \frac{  \log 2  }{\delta +  R }  = \frac{  \log 2  }{  \delta +   \sum_{k=1}^{8}  \exp \left( \epsilon_s \left\langle  \mathbf{s} , \mathbf{e}_k \right\rangle \right)   }  ,
\end{align*}
which proves maintext~\cref{eq:tmo_for_epsilon_n=0}.
Note that by the specific choice of our initial condition, which in particular uses the equilibrium fraction of unpolarised cells, there is no dependence on the parameter~$\beta$.
However, when comparing the effects of alignment dynamics to the effects of cell turnover it is natural to vary de novo polarisation rate~$\beta$ together with the death rate~$\delta$ to keep their ratio constant.
The time of minimal order $\tmoMF \left( \epsilon_n \right)$ decreases with increasing cell turnover, here apparent from $\delta$, and with increasing sensitivity $\epsilon_s$ to the global signal~$\mathbf{s}$.
For $\delta$ in the order of magnitude 1, the latter has the bigger impact on $\tmoMF \left( \epsilon_n \right)$ because $R \geq 8$ and $R$ grows exponentially with $\epsilon_s$.
In the limit of $\delta \to 0$, the largest time of minimal order is observed, yet it is still finite.
 
To further see the equality with the time of minimal order for the time series $\langle \mathbf{p} \left(t\right) \rangle$ claimed at the main text~\cref{eq:tmo_for_epsilon_n=0}, remember that the approximations in eqs.~\eqref{suppl:MFA_rate_pol_to_pol}, \eqref{suppl:MFA_rate_unpol_to_pol} and \eqref{suppl:MFA_rate_unpol_to_pol_shorter} are equalities for $\epsilon_n = 0$.
Hence the mean-field ODE \eqref{suppl:ODE_for_p_expressed_in_rR} in $\MFA{  \mathbf{p}  }$ is valid as well for $\langle \mathbf{p} \left(t\right) \rangle$, the order parameter in the IPS at time $t$ averaged across a sufficient number of realisations of the stochastic system.
Then the derivations shown above lead to an equation like \eqref{eq:tmo_for_epsilon_n=0} for the time of minimal order for the time series $\langle \mathbf{p} \left(t\right) \rangle$, finishing the proof of main text eq.~\eqref{eq:tmo_for_epsilon_n=0}.

\begin{figure}   %%% supplementary figure 5
{
  \includegraphics[width=1.0 \textwidth, keepaspectratio]{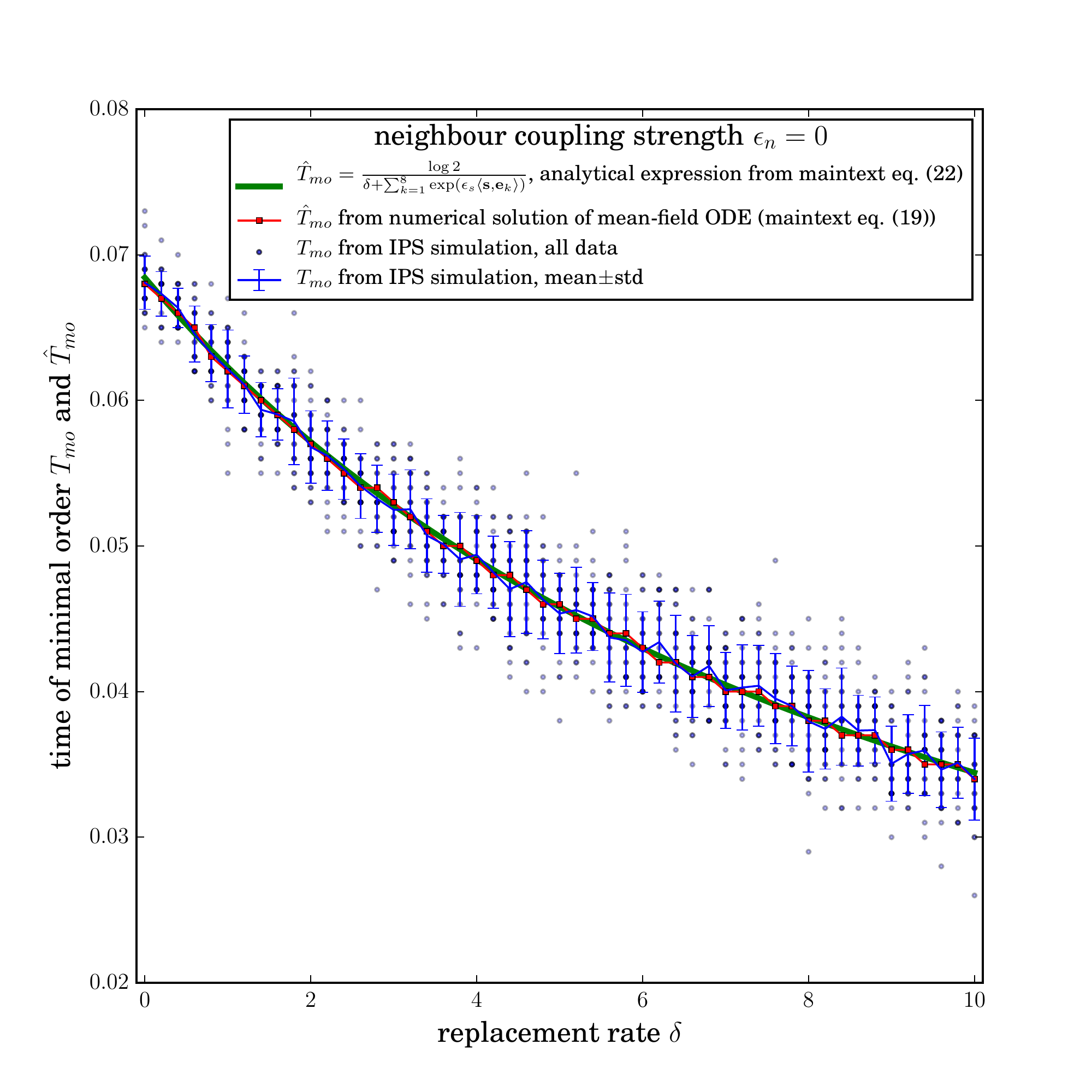} 
}

\caption[Time of minimal order for vanishing neighbour coupling strength~$\epsilon_n = 0$.]
{Comparison of time of minimal order obtained by different approaches for vanishing neighbour coupling strength~$\epsilon_n = 0$.
The time of minimal order $\tmoMF$ from numerical solutions of the mean-field ODE~\eqref{eq:ODE_expressed_in_rR} (red) coincides with the analytical expression given in main text eq.~\eqref{eq:tmo_for_epsilon_n=0} (green).
The time of minimal order $\tmo$  of the IPS scatter closely follows them. Average of 25~simulated trajectories (blue line) and all data points (blue transparent dots, 25 for each value of $\delta$). 
The transparent circles partly lay on top of each other because the state of the simulation was logged at a frequency of 0.001.
}
\label{fig:suppl:IPSvsMF_epsilon_n=0}
\end{figure}

Note that the time of minimal order for the mean order parameter~$\langle \mathbf{p} \left(t\right) \rangle$ might differ from the mean time of minimal order $ \langle \tmo \rangle$ in the IPS because the (in general non-linear) operator $\operatorname{argmin}_{t \in \left[0,1\right]}$ and averaging by $\langle \cdot \rangle$ are interchanged.
However, we observe close agreement between empirical $ \langle \tmo \rangle$ from 25~simulated trajectories of the IPS, the time of minimal order $\tmoMF$ from numerical solution of the mean-field model~\eqref{eq:ODE_expressed_in_rR} and the analytical expression of maintext~\cref{eq:tmo_for_epsilon_n=0} derived here, see suppl.~fig.~\ref{fig:suppl:IPSvsMF_epsilon_n=0}.

\section{Details on the numerical solution of the mean-field model} \label{SI:NumericalSolutionOfMean-fieldODE}  

This section extends maintext \cref{subsec:NumericalSolutionOfMean-fieldODE} by giving a more detailed description of the numerical solution of the mean-field model, see in particular maintext~\cref{eq:ODE_expressed_in_rR} and maintext~\cref{fig:results_Mean-field}.

The time courses of $ \left\| \MFA{ \mathbf{p} } \right\|$ and $\operatorname{ang} \left( \MFA{ \mathbf{p} } \right)$ for the turning cases exhibit several common characteristics independent of the specific parameter sets and with those for the original model, described as follows.
Throughout, $\MFA{a}_0 = \MFA{a}_0^{\ast} = \frac{\delta}{\beta + \delta}$, confirming the analytical prediction by eq.~\eqref{suppl:ODE_dead_fraction}.
According to the initialisation specified by eq.~\eqref{eq:initial_distribution}, the majority of polarised cells starts with polarisation direction $\mathbf{e}_4 = \left(-1, 0 \right)$, i.e. $\MFA{a}_4 \approx p_{eq} = \frac{\beta}{\beta + \delta}$, cf.~fig.~\ref{fig:results_Mean-field}A-C.
The fraction $\MFA{a}_4$ declines in favour of the other polarisation directions, in the beginning especially in favour of $\MFA{a}_3$ and $\MFA{a}_5$.
Then the fractions $\MFA{a}_2, \MFA{a}_6$ and to less extent $\MFA{a}_1, \MFA{a}_7$ and $\MFA{a}_8$ increase as well while $\MFA{a}_4$ declines further.
After $\MFA{a}_3, \MFA{a}_5$ start to decrease again, all fractions are approximately equally abundant at the time of minimal order $\tmoMF$.
Decline in $\MFA{a}_2$ to $\MFA{a}_6$ in favour of further increase in $\MFA{a}_1, \MFA{a}_7$ plus strong increase in $\MFA{a}_8$ leads into a plateau.
Because of symmetry, $\MFA{ \mathbf{p} } \left( t \right) = \left( \MFA{ p }_x \left( t \right), 0 \right)$, and especially $\MFA{ \mathbf{p} } \left(\tmoMF \right) = \left( 0, 0 \right) $.
The rates $r_k \left( \MFA{\mathbf{a}} \left( \tmoMF \right) \right)$ are biased towards $r_8$ because of the global signal, changing $\MFA{ p }_x$ from negative to positive sign and driving the polarity pattern towards a stable asymptotic state of dominant $\MFA{a}_8$ accompanied by major fractions $\MFA{a}_1, \MFA{a}_7$. 
In this asymptotic state, $\MFA{ \mathbf{p} }$ is parallel to $\mathbf{s} = \mathbf{e}_8$ and $\left\| \MFA{ \mathbf{p} } \right\| $ is almost as large as $p_{eq}$.
Because of coherence between global signal~$\mathbf{s}=\left(1,0 \right)$ and strong local signal $\MFA{ \mathbf{p} }$ in exactly the same direction, the solution reaches a stable equilibrium there.
The transition from alignment among cells conflicting with the global signal to alignment with the global signal happens via a disordered state when each polarisation direction is approximately equally abundant around $\tmoMF$, see fig.~\ref{fig:results_Mean-field}A-C.

The dynamics described are well recapitulated in the time course of the order parameter~$\MFA{ \mathbf{p} }$, see fig.~\ref{fig:results_Mean-field}C.
Modulus $\left\| \MFA{ \mathbf{p} } \right\| $ starts at a high value~$\approx \frac{\beta}{\beta + \delta}$, decreases to a distinct minimum $\left\| \MFA{ \mathbf{p} } \left(\tmoMF \right) \right\| = 0$ indicating complete disorder and then increases again to a plateau.  
The angle $\operatorname{ang} \left( \MFA{ \mathbf{p} } \right)$ first remains equal to~$- \pi$, and switches to~$0$ at $\tmoMF$.

\begin{figure}   %%% supplementary figure 4
{
  \includegraphics[width=1.0 \textwidth, keepaspectratio]{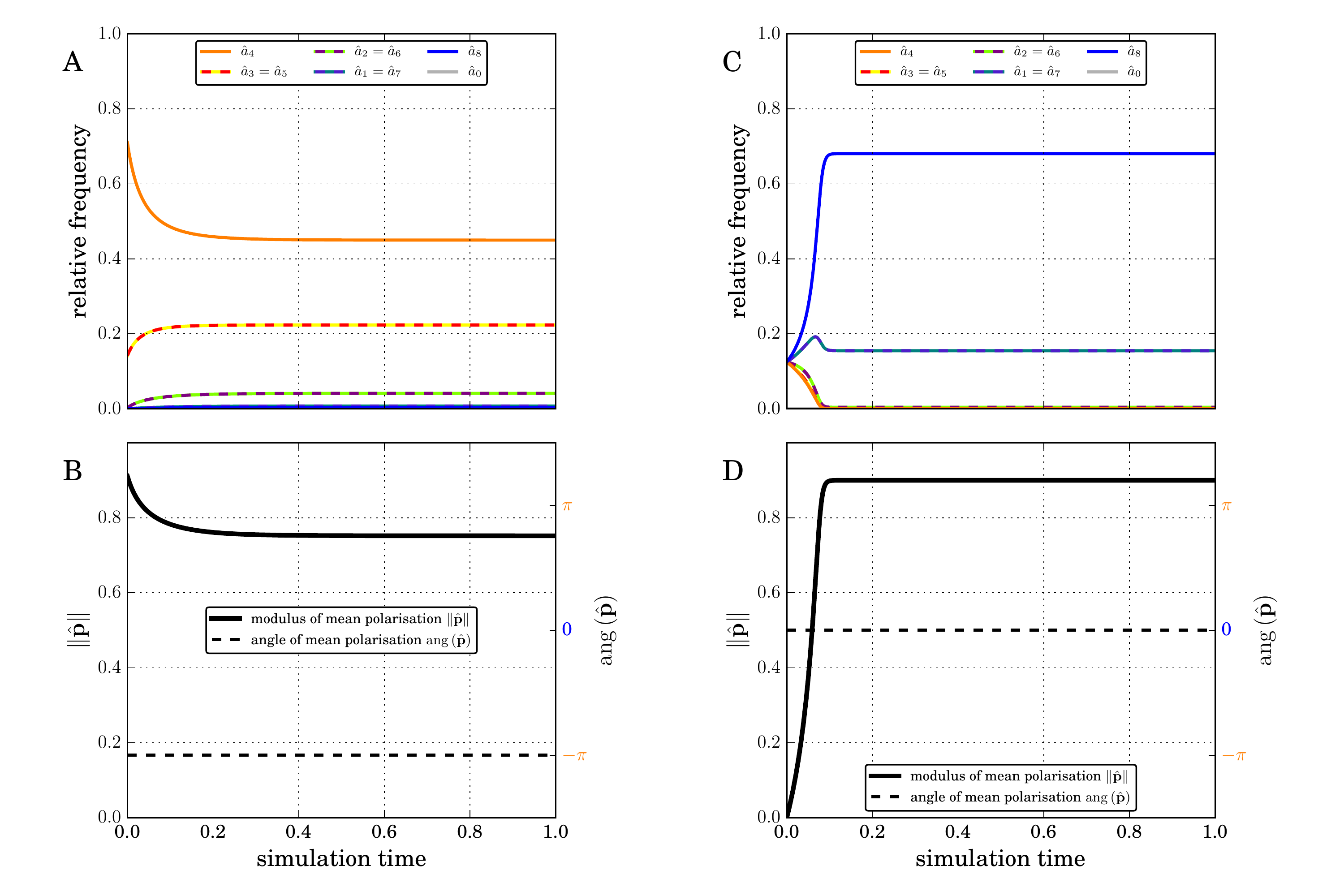} 
}
\caption[Bistability of the mean-field model ODE for very high neighbour coupling strength and low death rate.]
{The mean-field model ODE system~\eqref{eq:ODE_expressed_in_rR} is bistable for neighbour coupling strength $\epsilon_n > 4.5$ and cell death rate $\delta <  \delta_{crit} \left(\epsilon_n \right)$, exemplified here for $\epsilon_n = 4.5$, $\delta=0$, $\epsilon_s=1$, $\beta = 1$.
\textbf{A,B.} The polarity does not reverse when starting from the initial condition used throughout the paper, given by main text~eq.~\eqref{eq:initial_distribution}. 
Fraction $\MFA{a}_4$ decreases initially, but none of the fractions $\MFA{a}_8$, $\MFA{a}_1$, $\MFA{a}_7$ increases significantly.
Instead, the system approaches a stable steady state with high $\MFA{a}_4, \MFA{a}_3, \MFA{a}_5$ and $\MFA{ p }_x <0$, id est the mean polarisation $\MFA{ \mathbf{p}}$ remains pointing left \emph{counter-directional} to the global signal~$\mathbf{s}$. 
\textbf{C,D.} Using a different initial condition, namely $\MFA{a}_0 \left(0\right) = \frac{\delta}{\beta + \delta}$, $\MFA{a}_1 \left(0\right) = \mathellipsis = \MFA{a}_8 \left(0\right) = \frac{1}{8} \frac{\beta}{\beta + \delta}$, the system approaches a different stable steady state with $\MFA{ \mathbf{p}}$ aligned to the global signal~$\mathbf{s}$.
}
\label{fig:suppl:ODE_NoTurn}
\end{figure}

However, there is no time of minimal order at which $\left\| \MFA{ \mathbf{p} } \left(\tmoMF \right) \right\| = 0$ in the cases $\delta = 0, \, \epsilon_n \in \left\{4.5, 5 \right\}$ . 
Instead of approaching dominant $\MFA{a}_8, \MFA{a}_1, \MFA{a}_7$, the solution of the ODE system~\eqref{eq:ODE_expressed_in_rR} remains trapped in a stable asymptotic state with high $\MFA{a}_4, \MFA{a}_3, \MFA{a}_5$ and $\MFA{ p }_x <0$, see suppl. fig.~\ref{fig:suppl:ODE_NoTurn}A,B.  
Still a stable asymptotic state with dominant $\MFA{a}_8, \MFA{a}_1, \MFA{a}_7$ and $\MFA{ p }_x >0$ does exist, see suppl. fig.~\ref{fig:suppl:ODE_NoTurn}C,D, but the initial condition is not within its domain of attraction.
This trapping represents a phase transition to non-turning behaviour with diverging $\tmoMF$ as $\epsilon_n$ is increased and/or $\delta$ is decreased towards the critical parameter values.
However, this phase transition is only present in the mean-field approximation as the errors introduced by mean-field assumption grow with neighbour coupling strength~$\epsilon_n$, see main text sec.~\ref{subsubsec:mean-fieldAnalysis}. 
With increasing neighbour coupling, each single cell in the dynamically diluted alignment model is less probable to deviate from the initially dominant polarisation direction. 
However, such rare events of spontaneous polarity change still can occur in the original IPS and can initiate progressive polarity reorientation, whereas in the mean-field description the influence of deviating cells is neglected by averaging.
As increasing cell death rate $\delta$ reduces the expected fraction of polarised cells $p_{eq} = \frac{\beta}{\beta + \delta}$, cf. eq.~\eqref{eq:frac_pol}. 
This latter approximation introduces the phase transition into the mean-field model.
Note that a stable equilibrium of~\eqref{eq:ODE_expressed_in_rR} with dominant $a_8, a_1, a_7$ does still exist, 
but it is not reached from the initial state when another equilibrium with $\MFA{ p }_x <0$ arises for high neighbour coupling~$\epsilon_n$, cf. suppl. fig.~\ref{fig:suppl:ODE_NoTurn}A,B versus C,D. 
Close to that transition and beyond, the mean-field description is no longer a valid approximation of the IPS model.
Therefore we focus the discussion in the main text on the parameter range of lower $\epsilon_n$ and/or larger $\delta$.

\newpage

\begin{figure}   %%% supplementary figure 1
{
  \includegraphics[width=1.0 \textwidth, keepaspectratio]{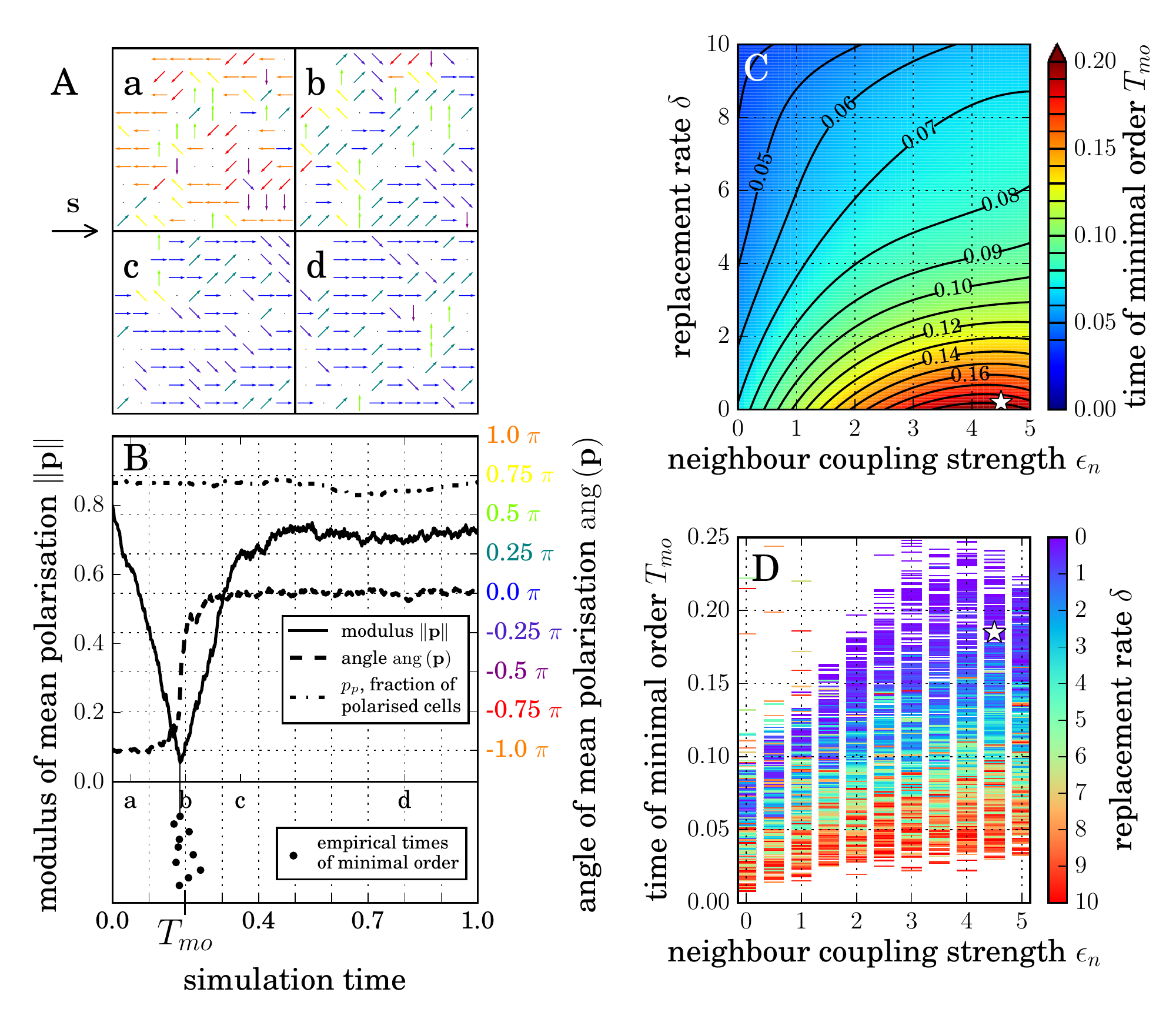} 
}
\caption[Data from IPS simulations for lattice size of $20 \times 20$.]
{
Supplement to maintext fig.~\ref{fig:Results_IPS_size100}.
Data shown here is analogous to that in maintext fig.~\ref{fig:Results_IPS_size100}, except for lattice size of $20 \times 20$ used here.
Qualitative and quantitative similarity indicates that lattice size of $100 \times 100$ employed in the maintext is sufficient to avoid finite size effects. 
\textbf{A.} Snapshots of a typical simulation at times~0.05,~0.2,~0.5,~0.8 (\textbf{a-d}), see fig.~\ref{fig:ModelDef}C for colour code. 
Zoomed details of a $20 \times 20$~lattice with parameters as in maintext fig.~\ref{fig:Results_IPS_size100}A, id est periodic boundaries, $\epsilon_n=4.5$, $\delta=0.2$, $\beta=1$, $\epsilon_s=1$, $\mathbf{s}=\left(1,0\right)$. 
\textbf{B.} Mean polarisation vector~$\mathbf{p}$ depicted as modulus~$\left\| \mathbf{p} \right\| $ (solid, left axis) and angle~$\operatorname{ang} \left( \mathbf{p} \right)$ (dashed, right axis, colour code as in panel A and fig~\ref{fig:ModelDef}C). 
The distinctive minimum of the $\left\| \mathbf{p} \right\|$ time course defines the time of minimal order~$\tmo$.
The fraction of polarised cells $p_p$ (dash-dotted, left axis) fluctuates around $p_{eq} = 0.8{\bar{3}}$.
Fluctuations in $\left\| \mathbf{p} \right\| $, $\operatorname{ang} \left( \mathbf{p} \right)$ and $p_p$ are stronger than in maintext fig.~\ref{fig:Results_IPS_size100}B, and the $T_{mo,i}$ from time courses $i \in \mathbb{N}$ for equal parameters are more scattered.
Nevertheless, characteristics of the time courses are preserved, where $100 \times 100$ lattice yields even smoother trajectories.
\textbf{C,D.} Simulation results for mean $\tmo$ of 25 repetitions shown as heatmap with contourlines (isotemporales at marked levels, \textbf{C}) and all data points (\textbf{D}), fixed parameters as in A,B and maintext fig.~\ref{fig:Results_IPS_size100}A,B. Asterisks denote parameter values of panels A,B and maintext fig.~\ref{fig:Results_IPS_size100}A,B. For the mean $\tmo$, differences between $20 \times 20$ and $100 \times 100$ lattice (C versus maintext fig.~\ref{fig:Results_IPS_size100}C) are marginal. Note inverted colour bar for $\delta$ in \textbf{D}.
}
\label{fig:suppl:Results_IPS_size20}
\end{figure}

\newpage

\begin{figure}   %%% supplementary figure 2
{
  \includegraphics[width=1.0 \textwidth, keepaspectratio]{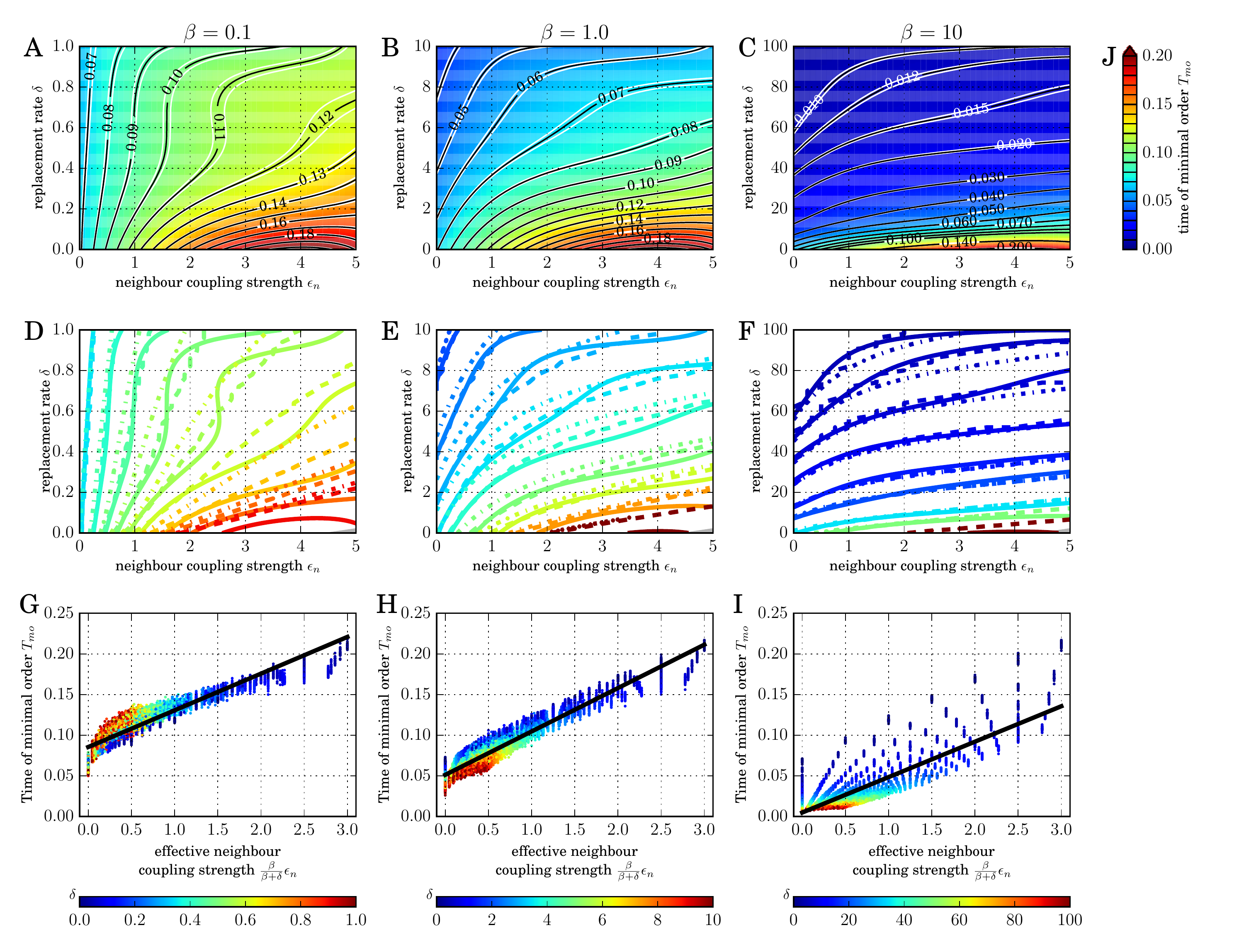} 
}
\caption[Effects of de novo polarisation rate $\beta$ on the time of minimal order $\tmo$. ]
{
Supplement to maintext fig.~\ref{fig:Results_IPS_size100}.
Effects of de novo polarisation rate $\beta$ on the time of minimal order $\tmo$. 
The time of minimal order $\tmo$ in IPS simulations is statistically robust across the parameter space, and measured data collapse onto a linear dependence upon rescaling to \textit{effective} neighbour coupling strength. 
\textbf{A-C.} Time of minimal order for $\beta = 0.1$(\textbf{A}), $\beta = 1$(\textbf{B}), and $\beta = 10$(\textbf{C}) with lattice size $100 \times 100$, $\epsilon_s=1$, $\mathbf{s} = \left( 1, 0 \right)$. Heatmap and black isotemporales at marked levels are for mean, surrounding white isotemporales for $\textnormal{mean} \pm \textnormal{sem}$ from $25$~repetitions for each data point.
Panel B is a reproduction of maintext fig.~\ref{fig:Results_IPS_size100}C with white isotemporales added.
Colour code is common to A-F and the same as in maintext fig.~\ref{fig:Results_IPS_size100}C.
\textbf{D-F.} Comparison of isotemporales for IPS model (solid),  mean-field model (dashed) and linearised mean-field model (dash-dot).
Shown levels are $0.07, 0.08, 0.09, 0.10, 0.11, 0.12, 0.14, 0.16, 0.18$ (\textbf{D}, $\beta = 0.1$) and $0.04, 0.05, 0.06, 0.07, 0.08, 0.10, 0.12, 0.15, 0.20$ (\textbf{E}, $\beta = 1$) and $0.01, 0.012, 0.015, 0.02, 0.03, 0.04, 0.07, 0.10, 0.20$ (\textbf{F}, $\beta = 10$), respectively.
For $\epsilon_n > 4.5$ the mean-field model undergoes a phase transition to non-turning dynamics (grey shaded region in bottom right of D-F) for all values $\beta = 0.1, 1, 10$ studied, see also maintext~\cref{fig:results_Mean-field}E.
\textbf{G-I.} All data of IPS simulation for the respective value of $\beta$ (A-C) collapse to approximately linear dependence (black curves).
Replacement rate $\delta$ is colour-coded to have fixed ratio with $\beta = 0.1$ (\textbf{G}), $\beta = 1$ (\textbf{H}) and $\beta = 10$ (\textbf{I}), respectively (horizontal colour-bars). 
Panel H is equivalent to maintext~\cref{fig:Result_TimeOfMinimalOrder_in_IPS_collapse}B.
Additionally accounting for the time offset $\tmo \left(\epsilon_n =0\right)$ from maintext~\cref{eq:tmo_for_epsilon_n=0} reduces the scatter further, shown in \cref{fig:suppl:Result_TimeOfMinimalOrder_in_IPS_collapse_further} for $\beta =1$. 
\textbf{J} Colour-code for time of minimal order $\tmo$ and $\tmoMF$ in A-F.
}
\label{fig:suppl:Results_IPS_size100_beta0_1and1and10}
\end{figure}

\newpage

\begin{figure}   %%% supplementary figure 3
{
  \includegraphics[width=1.0 \textwidth, keepaspectratio]{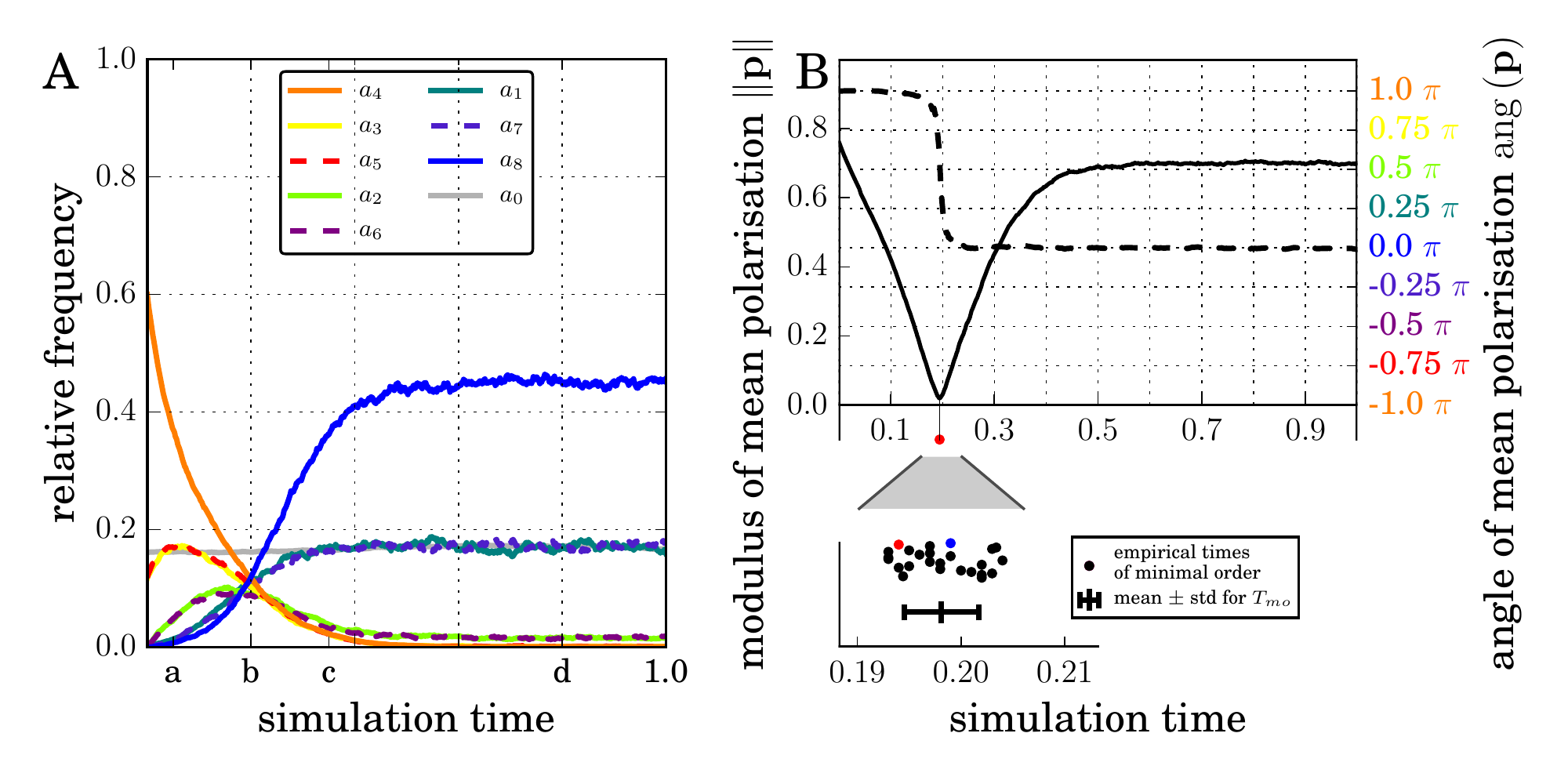} 
}
\caption[Fractions $a_0, a_1, \mathellipsis, a_8  $ for example trajectory; statistic robustness of time of minimal order $\tmo$.]
{
Supplement to maintext~\cref{fig:Results_IPS_size100}.
\textbf{A.} Fractions $a_0, a_1, \mathellipsis, a_8  $ of nodes in state $\mathbf{e}_0, \mathbf{e}_1, \mathellipsis, \mathbf{e}_8$, respectively, of the example simulation shown in fig.~\ref{fig:Results_IPS_size100}A,B and suppl.~movie~2.
Parameters are $\epsilon_n=4.5$, $\delta=0.2$, $\beta=1$, $\epsilon_s=1$, $\mathbf{s}=\left(1,0\right)$. 
Observe high agreement with $\MFA{a}_0, \MFA{a}_1, \mathellipsis, \MFA{a}_8  $ in mean-field approximation, see fig.~\ref{fig:results_Mean-field}B using the same parameters.
\textbf{B.} The time of minimal order $\tmo$ in IPS simulations is statistically robust for a fixed parameter set.
Upper part: Example simulation with turn of mean polarisation vector clockwise, compared to counter-clockwise turn shown in maintext fig.~\ref{fig:Results_IPS_size100}B. 
Modulus solid, angle dashed.
Note the starting angle of $\pi$ is equivalent to $-\pi$ shown in maintext~\cref{fig:Results_IPS_size100}B as angular argument.
Parameters as in fig.~\ref{fig:Results_IPS_size100}A,B and in suppl.~movie~2 ($\epsilon_n=4.5$, $\delta=0.2$, $\beta=1$).
Lower part: statistics of $\tmo$ from 25~sampled trajectories for the same parameter set.
Red and blue dots indicate the time of minimal order of the trajectories shown in the upper part and in maintext~\cref{fig:Results_IPS_size100}A,B, respectively. 
}
\label{fig:suppl:Results_IPS_size100_beta1_statistics}
\end{figure}

\newpage

\newpage

\newpage

\begin{figure}   %%% supplementary figure 6
{
  \includegraphics[width=1.0 \textwidth, keepaspectratio]{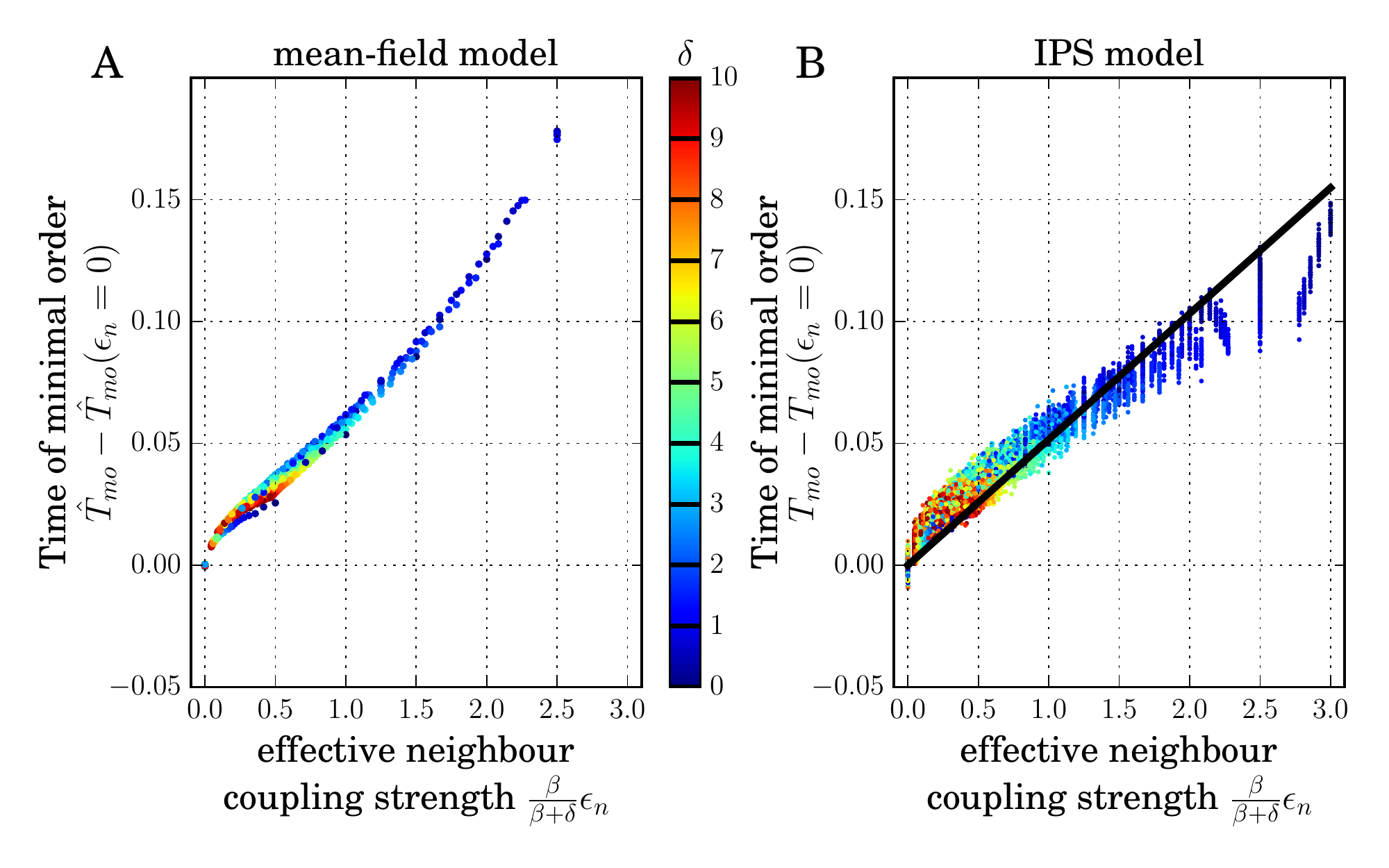} 
}
\caption[Time of minimal order $\tmo$ collapses to linear function of effective neighbour coupling strength $\epsilon_n^{eff}$]
{
Supplement to maintext~\cref{fig:Result_TimeOfMinimalOrder_in_IPS_collapse}, where measured data collapse onto a linear dependence upon rescaling to \textit{effective} neighbour coupling strength, cf. maintext~\cref{eq:tmo=functionOf_eps_n}.
The small remaining scatter of the data points around the black line in maintext~\cref{fig:Result_TimeOfMinimalOrder_in_IPS_collapse}B is largely given by a $\delta$-dependent offset as the ordered colours of the data points indicate.
This $\delta$-dependent offset is analytically known for $\epsilon_n=0$ by maintext~\cref{eq:tmo_for_epsilon_n=0}.
Subtracting the offset at  $\epsilon_n=0$, where $\tmo \left( \epsilon_n = 0 \right) = \tmoMF \left( \epsilon_n =0 \right) = \frac{\log 2}{ \delta + \sum_{k=1}^{8} \exp \left( \epsilon_s \left\langle  \mathbf{s}, \mathbf{e}_k \right\rangle  \right) }$ is taken from the analytical expression in maintext~\cref{eq:tmo_for_epsilon_n=0}, reduces the scatter further.
\textbf{A.} All data of the mean-field model, $\beta = 1$.
Colour code of cell death rate~$\delta$ applies to both panels.
\textbf{B.} All data of the IPS simulations, $\beta = 1$.
The black line obeys 

$\qquad \qquad  \tmo = \tmo \left( \epsilon_n^{eff} \right) = (0.0516\pm0.0001) \underbrace{  \frac{\beta}{\beta+\delta}  }_{ =p_{eq} } \epsilon_n +  \tmo \left( \epsilon_n = 0 \right)$

where values in brackets denote $(\textnormal{mean} \pm \textnormal{std} )$ of an orthogonal distance regression.
}
\label{fig:suppl:Result_TimeOfMinimalOrder_in_IPS_collapse_further}
\end{figure}

\newpage

\begin{figure}   %%% supplementary figure 7
{
  \includegraphics[width=1.0 \textwidth, keepaspectratio]{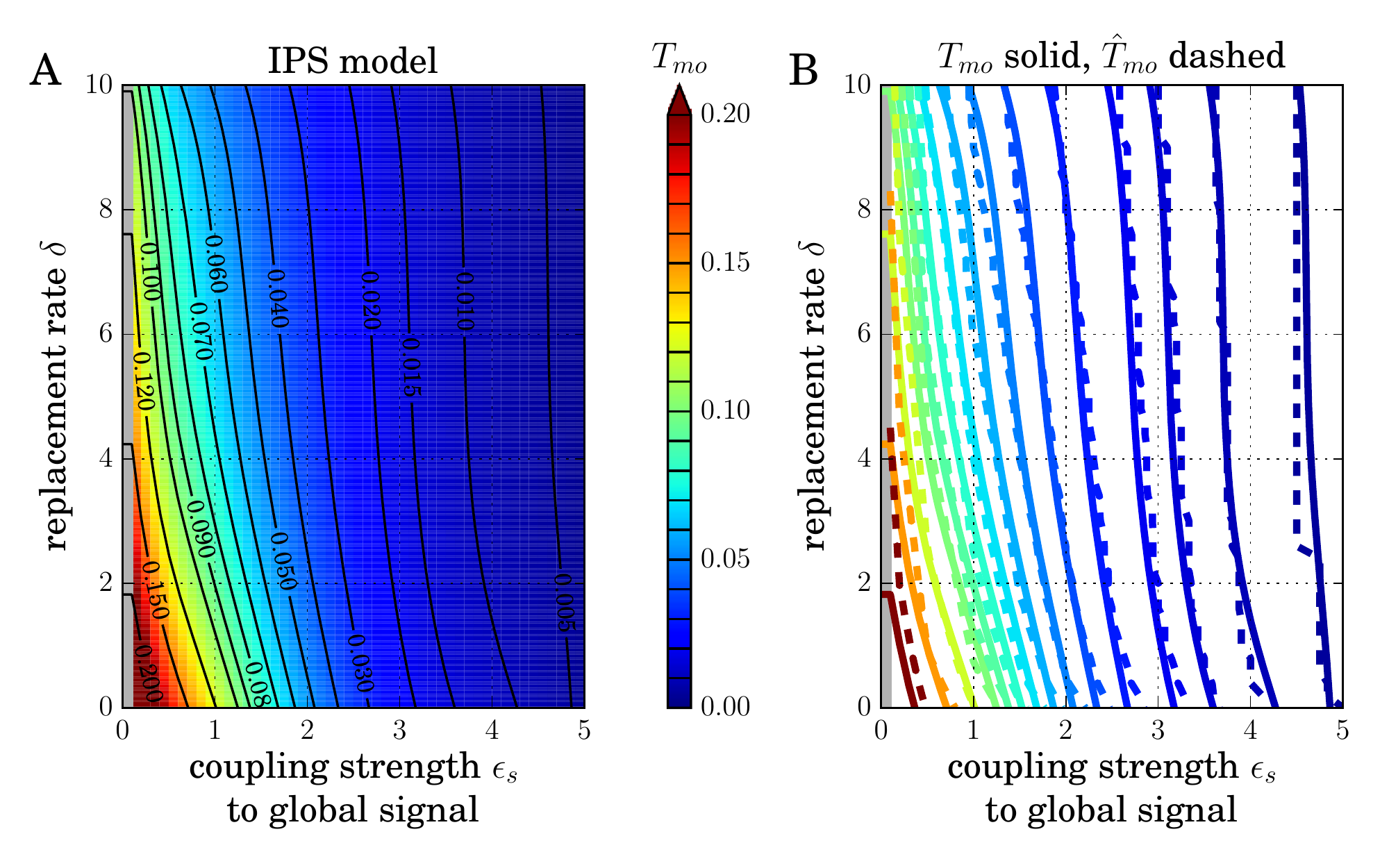} 
}
\caption[Dependence of time of minimal order $\tmo$ on the coupling strength $\epsilon_s$ to the global signal]
{
The time of minimal order $\tmo$ depends smoothly on the coupling strength $\epsilon_s$ to the global signal, and IPS and mean-field model agree closely.
\textbf{A.} IPS simulation results for mean $\tmo$ of 25 repetitions shown as heatmap with contourlines (isotemporales at marked levels). 
\textbf{B.} Comparison of time of minimal order as contourlines for IPS model ($\tmo$, solid) and mean-field model ($\tmoMF$, dashed).
Colour code of the time of minimal order applies to both panels.
For all values $\epsilon_s \in \left\{ 0.1, \, 0.2, \, 0.5, \, 1.0, \, 2.0, \, 3.0, \, 4.0, \, 5.0 \right\}$ the tissue polarity pattern reorganises, both in the IPS simulations and in the mean-field model.
The time of minimal order decreases for higher coupling strength $\epsilon_s$ to the global signal, as predicted. 
For $\epsilon_s = 0.0$, there is no influence of a global signal and no dominant polarisation direction develops (grey shaded areas in A,B).
Note that by the choice of our initial condition, cf. maintext~\cref{eq:initial_distribution}, no dominant polarisation direction exists for $\epsilon_s = 0.0$ in the initial state either.
Other parameters $\epsilon_n=1.0$, $\beta=1.0$.
}
\label{fig:suppl:Result_TimeOfMinimalOrder_as_function_of_epsilon_s}
\end{figure}

\end{document}